%% file: main.tex
\documentclass[fleqn, usenatbib]{mnras}

\usepackage{newtxtext,newtxmath}
\usepackage[utf8]{inputenc}
\usepackage[table]{xcolor}
\usepackage{hyperref}
\usepackage{subfigure}
\usepackage{mathtools}
\usepackage{geometry}
\usepackage{comment}
\usepackage[section]{placeins}
\usepackage{booktabs}
\usepackage{float}
\usepackage{cleveref}
\usepackage{glossaries}
\usepackage{physics}
\usepackage{float}
\usepackage[english]{babel}
\usepackage{blindtext}
\usepackage{tikz}
\usetikzlibrary{arrows.meta}
\usepackage{soul}
\usepackage[T1]{fontenc}
\usepackage{ae,aecompl}
\usepackage{graphicx}	
\usepackage{ulem}

\newcommand{\pd}{\partial}

\definecolor{light-gray}{gray}{0.95}
\definecolor{greeen}{rgb}{0 0.5 0}
\newcommand{\code}[1]{\colorbox{light-gray}{\textsc{#1}}}

\title[Hadron-Quark Phase Transition in Supernovae]{The Role of the Hadron-Quark Phase Transition in Core-Collapse Supernovae} 
\author[P. Jakobus et al.]{
Pia~Jakobus$^{1,2}$\thanks{E-mail: pia.jakobus@monash.edu},
Bernhard~M\"uller$^{1,2}$,
Alexander~Heger$^{1,2,3,4}$,
Anton Motornenko$^{5}$,
\newauthor Jan Steinheimer$^{5}$, 
Horst Stoecker$^{5,6,7}$
\\
$^{1}$ School of Physics and Astronomy, Monash University, VIC 3800, Australia\\
$^{2}$ Australian Research Council Centre of Excellence for Gravitational Wave Discovery (OzGrav), Clayton, VIC 3800, Australia \\
$^{3}$ Center of Excellence for Astrophysics in Three Dimensions (ASTRO-3D), Australia\\
$^{4}$ The Joint Institute for Nuclear Astrophysics, Michigan State University, East Lansing, Michigan 48824, USA \\
$^{5}$ Frankfurt Institute for Advanced Studies, Giersch Science Center, Frankfurt am Main, Germany\\
$^{6}$ Institut für Theoretische Physik, Goethe Universität, Frankfurt am Main, Germany \\
$^{7}$ GSI Helmholtzzentrum für Schwerionenforschung GmbH, Darmstadt, Germany}
\date{Accepted XXX. Received YYY; in original form ZZZ}

\makeglossaries
\newacronym{qse}{QSE}{Quasi-Statistical equilibrium}
\newacronym{nse}{NSE}{Nuclear-Statistical equilibrium}
\newacronym{qcd}{QCD}{Quantum Chromodynamics}
\newacronym{bsg}{BSG}{Blue supergiant stars}

\newacronym{zams}{ZAMS}{zero-age main sequence}
\newacronym{qgp}{QGP}{quark-gluon plasma}
\newacronym{ccs}{CCSN}{core-collapse supernovae}
\newacronym{rmf}{RMF}{relativistic mean-field}
\newacronym{tov}{TOV}{Tolman-Oppenheimer-Volkoff} %\glspl is plural
\newacronym{eos}{EoS}{equation of state}
\newacronym[longplural={mass radius relations}]{mr}{M-R relation}{mass radius relation}

\begin{document}
\label{firstpage}
\pagerange{\pageref{firstpage}--\pageref{lastpage}}
\maketitle
\begin{abstract}
The hadron-quark phase transition in quantum chromodyanmics has been suggested as an alternative explosion mechanism for core-collapse supernovae. We study the impact of three different hadron-quark equations of state (EoS) with first-order (DD2F\_SF, STOS-B145) and second-order (CMF) phase transitions on supernova dynamics by performing 97 simulations for solar- and zero-metallicity progenitors in the range of $14\texttt{-}100\,\text{M}_\odot$. We find explosions only for two low-compactness models ($14 \text{M}_\odot$ and $16\,\text{M}_\odot$) with the DD2F\_SF EoS, both with low explosion energies of $\mathord{\sim}10^{50}\,\mathrm{erg}$. These weak explosions are characterised by a neutrino signal with several mini-bursts in the explosion phase due to complex reverse shock dynamics, in addition to the typical second neutrino burst for phase-transition driven explosions. The nucleosynthesis shows significant overproduction of nuclei such as $^{90}\mathrm{Zr}$ for the $14\,\text{M}_\odot$ zero-metallicity model and $^{94}\mathrm{Zr}$ for the $16\,\text{M}_\odot$  solar-metallicity model, but the overproduction factors are not large enough to place constraints on the occurrence of such explosions.  Several other low-compactness models using the DD2F\_SF EoS and two high-compactness models using the STOS EoS end up as failed explosions and emit a second neutrino burst.  For the CMF EoS, the phase transition never leads to a second bounce and explosion. For all three EoS, inverted convection occurs deep in the core of the proto-compact star due to anomalous behaviour of thermodynamic derivatives in the mixed phase, which heats the core to entropies up to $4k_\text{B}/\text{baryon}$ and may have a distinctive gravitational wave signature, also for a second-order phase transition. 
\end{abstract}
\section{Introduction}
The iron core of massive stars will undergo gravitational collapse once its mass exceeds the effective Chandrasekhar limit. The minimum mass for stars to burn their cores up to iron is estimated to be about $8 M_\odot$ \citep{woosley_02,ibeling_13}. 
The reduction of degeneracy pressure by electron captures and
photodisintegration of heavy nuclei eventually trigger runaway collapse on a free-fall time scale.
The core density increases up to the point where repulsive short-range nuclear interactions come into play, leading to a stiffening of the equation of state and a (first) core bounce, leaving a proto-neutron star -- or more generally a ``proto-compact star'' (PCS) if non-nucleonic particles eventually appear -- at the centre of the star.
The rebound of the core launches a shock wave that quickly stalls as its initial kinetic energy is drained by photodissociation of heavy nuclei and neutrino losses as it propagates through the outer core.
The subsequent evolution of the shock wave and the high-density EoS determine whether the star ends its life as a compact remnant in form of either a  black hole (BH), neutron star (NS),  quark star or some other compact star containing non-nucleonic matter.
Several mechanisms may revive the shock and result in a
core-collapse supernova (CCSN) explosion. The two best-explored scenarios are neutrino-driven explosions, which probably account for most explosions of up to $\mathord{\sim}10^{51}\,\mathrm{erg}$,  and the magnetorotational  mechanism, which has been proposed to explain exceptionally energetic ``hypernovae'' with explosion energies of up to $\mathord{\sim}10^{52}\,\mathrm{erg}$ 
\citep[for reviews, see][]{janka_12,burrows_13,mueller_20}. Both of these mechanisms critically rely on multi-dimensional effects such as rotation, convection, and other hydrodynamic instabilities.
A third proposed mechanism is the phase-transition driven (PT-driven) mechanism~\citep{MIGDAL1979158}, in particular a transition from hadrons to quarks~\citep{Sagert2009-nd,Fischer2017lag}. 

By and large, supernova matter requires a conformable treatment for numerous intrinsically different thermodynamic regimes. At temperatures below $\approx 6$ GK the baryon EoS has to account for heavy nuclei and their abundances, which are usually not in equilibrium with each other and need to be calculated by a  nuclear reaction network. At higher temperatures and later stages of the evolution, nuclear statistical equilibrium (NSE) can be applied. From the collapse phase onward, nuclear interactions need to be modelled explicitly at high densities instead of assuming non-interacting nucleons and nuclei. 
While most current CCSN simulations consider purely nucleonic high-density EoSs (with popular choices including \citealt{Shen1998-jc, Shen1998-hy, Lattimer1991-li,Hempel:2009mc,steiner_13,2012ApJ...748...70H}), the state of matter is highly uncertain at temperatures of several $10\, \mathrm{GK}$ and densities well exceeding the nuclear saturation density $n_0$. It is possible that quarks exist in supernova matter~\citep{Gentile1993-ss,Bednarek:1996nd,Drago_1999}; this is  not considered in most standard CCSN models. 

\begin{figure}
    \centering
    \includegraphics[width=\linewidth]{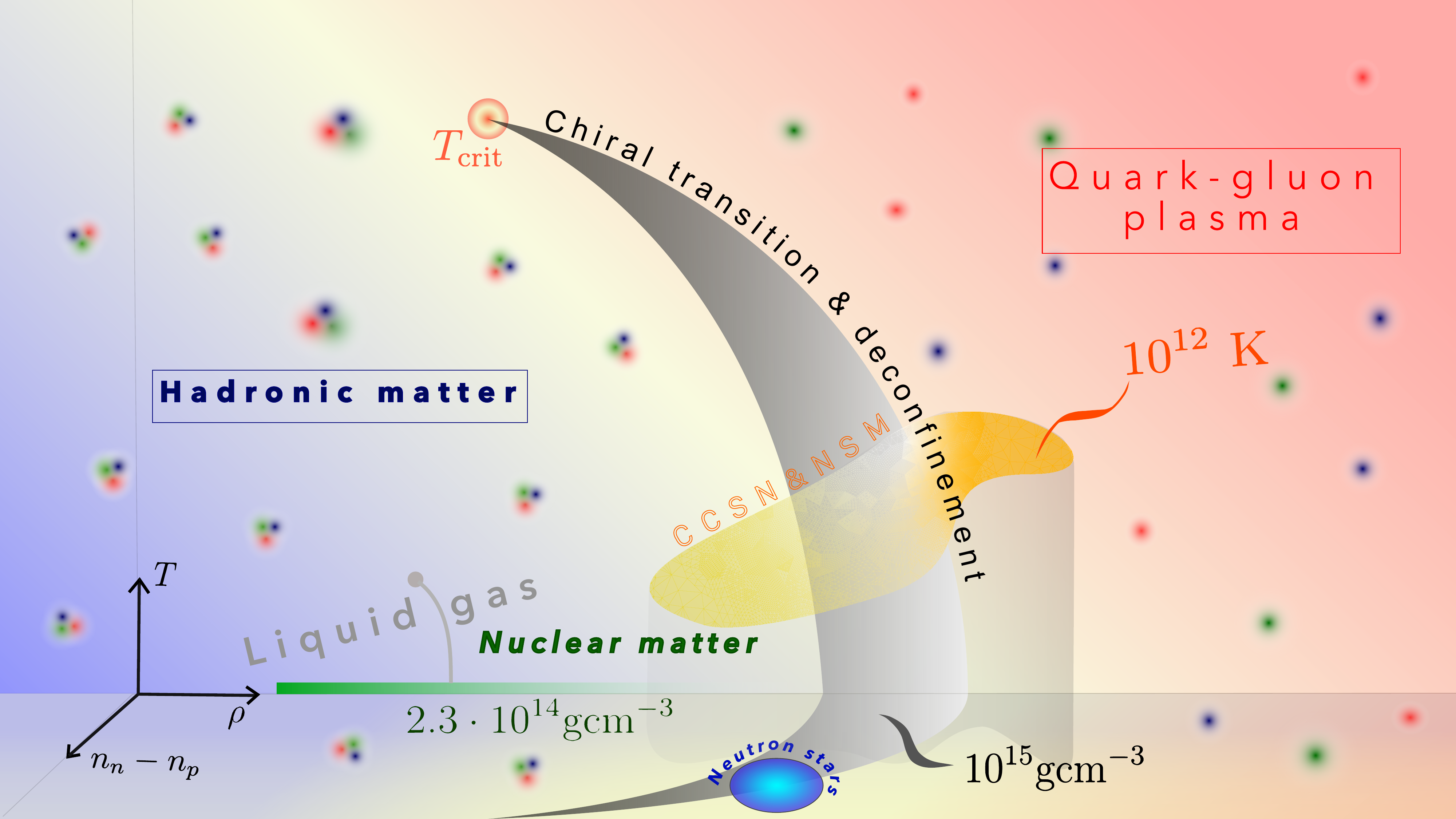}
    \caption{Illustrative QCD phase diagram. The $x$-, $y$-, and $z$-axis represent density, temperature and isospin asymmetry density $n_n-n_p$. The regime of late post-bounce phase of core-collapse supernovae and of neutron-star mergers are indicated by the orange band and grey shade.}
    \label{fig:phased}
\end{figure}

The relevant degrees of freedom in the high-density and high-temperature phase of the QCD phase diagram are free quarks and gluons~\citep{1984PhRvD..30..272W}. Perturbative methods in this regime give expressions for the thermodynamical potential $\Omega$ in powers of the QCD strong coupling constant $\alpha_s$. The latter is density- and temperature-dependent, and becomes small when the quark chemical potential is high or the temperature is large compared to the QCD energy scale $\Lambda_\text{QCD}$, which corresponds to large chemical potentials of several GeV~\citep{Baym2018-ti}. This is called asymptotic freedom;  quarks and gluons in this regime barely interact and can move freely~\citep{Peskin2019_my}. This regime is beyond the densities reached in CCSN. Regions of the QCD phase diagram with lower temperatures and chemical potentials prohibit the use of perturbative methods since the the running coupling $\alpha_\mathrm{s}$ becomes large and the series expansion of the thermodynamical potential $\Omega$ does not converge. Lattice QCD (lQCD) calculations provide insight into matter at near-zero densities and finite temperature~\citep{Gattringer2010_1,Gattringer2010_2}. 
The solution for QCD predicts a smooth crossover transition near vanishing baryon density at a pseudo-critical temperature in the range $150\texttt{-}160\,\mathrm{MeV}$~\citep{PhysRevD.90.094503, 201915,BORSANYI201499}. Several decades of high-energy heavy-ion collision analysis further advanced our understanding on QCD matter~\citep{stoecker_1986,gazdzicki_1998,Csernai_2013,Vovchenko_2019}. Heavy-ion collision data indicate a transition from bound hadrons to deconfined quarks at non-vanishing baryon density~\citep{adams_2005,arsene_2005,Adamczyk_2017}. The order of the phase transition at finite densities is, however, not yet fully understood \citep{osti_1774074,Cuteri:2021ikv}. 
With the advancement of multi-messenger astronomy~\citep{Abbott_2017,Aartsen_2017}, astrophysical observations have the potential to further constrain the EoS in the QCD regime as the physical conditions in relativistic heavy-ion collisions are closely linked, e.g., to those in neutron star mergers~\citep{musch_2019,hanauske_2017,hanauske_2019,most_2022}. 
The extreme environment at the centre of PCSs can exceed densities of $\rho_\text{centre} \gtrsim 10^{15}\, \mathrm{g}\,\mathrm{cm}^{-3}$ with maximum temperatures of the order $10^{12}\,\mathrm{K}$.  
Phenomenological models are the basis on which predictions about the phase transition from hadrons to quarks rely in these intermediate regimes of the QCD phase diagram. 
An illustration of the different regimes is shown in Figure~\ref{fig:phased}.\\
The hadron-quark phase transition usually softens the EoS during the mixed phase, which lowers the maximal supported mass~$M_\text{max}$ of the PCS. When the PCS exceeds $M_\text{max}$, it collapses and can perform damped oscillations around its new equilibrium position. If rapid BH formation does immediately follow, the core can bounce a second time, which can, depending on the released binding energy, lead to the formation of a shock wave, which then again can lead to expelling outer layers of the star, leaving behind a stable compact star \citep{Sagert2009-nd,Fischer2017lag}.

It is assumed that less massive progenitors with zero-age main sequence (ZAMS) mass
$\mathord{\lesssim} 20$~M$_\odot$ explode by the standard NDE-mechanism~\citep{nakamura_15,sukhbold_16,mueller_16a,burrows_20}, leaving behind NSs, while more massive stars usually collapse into BHs, even if shock revival may sometimes occur and be followed by a weak fallback explosion
\citep[e.g.,][]{chan_18,chan_20,ott_18,10.1093/mnrasl/sly059,powell_21,rahman_21}. So far, three-dimensional models of neutrino-driven explosions can only account for CCSNe of normal energies, not significantly exceeding $10^{51}\, \mathrm{erg}$ \citep{burrows_20,powell_20,bollig_21}.
A different mechanism for the most energetic observed CCSNe is probably needed. While the magnetorotational mechanism has long been investigated as an explanation for the most powerful explosions (for recent results on explosion energies, see \citealp{kuroda_20,obergaulinger_21,jardine_22}), 
recent studies discussed PT-driven explosions of very massive progenitors as a scenario for various unusually energetic events.
For a $50 M_\odot$ progenitor, \citet{Fischer2017lag} obtained an explosion energy of $E_\text{exp} = 3\times 10^{51}\,\mathrm{erg}$ in spherical symmetry.
The possibility of PT-driven explosions is also relevant for nucleosynthesis. Successful PT-driven explosions have been proposed a site for heavy  $r$-process elements \citep{Nishimura2011-yo,Fischer2020-rg}.  Furthermore,
a Galactic PT-supernova would be a promising target for
multi-messenger observations in neutrinos and gravitational waves. The neutrino signal would provide a characteristic fingerprint for the  QCD phase-transition in the form
of an electron \emph{anti}neutrino burst \citep{Sagert2009-nd,Fischer2017lag,Zha2020-rv} that would clearly be observable by present and future detectors such as IceCube, Super-Kamiokande, and Hyper-K. 
A first-order QCD phase transition is also expected to produce a strong and characteristic gravitational wave signal peaking at several kHz, regardless of whether a successful explosion ensues (\citealp{Zha2020-rv,Kuroda2021-zc}; see also
\citealp{osti_21020381} for other effects on the gravitational wave signal).

The robustness of the PT-driven mechanism, however, is far from clear yet. At this stage, it is important to more systematically scan the parameter space for PT-driven explosions using larger sets of progenitors 
($\mathord{\gg} 20$) than in the currently available literature. Furthermore only first-order phase transitions were considered  so far. Recent studies motivate further investigation into the robustness of the PT-driven scenario.
\citet{Zha2021-xk} found no successful explosions in 1D simulations using the STOS-B145 EoS, although they found instances of second bounces to a more compact and (transiently) stable PCS, with a strong dependence of the dynamics on the compactness parameter. Using the DD2F\_SF EoS, \citet{Fischer2021-zh} recently found explosions only at solar metallicity, but not at low metallicity for two $75 M_\odot$ models. 

This paper aims to shed more light on the progenitor dependence of the post-bounce evolution of hybrid PCSs containing quark matter. We perform 97 general-relativistic hydrodynamic simulations with neutrino transport
in spherical symmetry for up to 40 progenitors in the mass range $M\in [14,100] M_\odot$ with solar and zero metallicity using three different hybrid EoS. We especially focus on the thermodynamic features of the mixed phase and crossover regions and how these affect the post-bounce dynamics once the PCS reaches the threshold density for the appearance of quarks. We find that only two models using  the DD2F\_SF-1.4 EoS explode. For those two cases we perform a detailed nucleosynthesis analysis. We also study the neutrino signals of exploding and non-exploding models confirming a similar phenomenology as found by \citet{Zha2020-rv}.

This work is structured as follows. In Section~\ref{section two} we discuss the set of equations of state and give an overview of numerical methods including progenitor setup and the nucleosynthesis post-processing. In Section~\ref{section three} we present the results of our simulations. We interpret the progenitor dependence based on the DD2F\_SF EOS and discuss detailed hydrodynamic post-bounce dynamics for two exploding models in the DD2F\_SF setup. We analyse the effect of phase transitions by comparing hydrodynamic simulation outcomes for different EoS, including the neutrino signals.  Lastly, we review nucleosynthesis results in the ejecta for two exploding models. We summarise and our findings and their implications in Section~\ref{section four}.

\section{Simulation Setup}\label{section two}

\subsection{Equations of State with Quark Matter}
We study the effect of a quark-hadron phase transition in CCSNe using three high-density EoS with different treatments for the hadronic phase, the quark phase, and the phase transition between them.

\subsubsection{DD2-RMF EoS}
The DD2F\_SF EoS is a hadron-quark EoS featuring a first-order phase transition to deconfined quark matter. The EoS
belongs to a new class of hybrid EoS, using a relativistic density-functional formalism~\citep{Fischer2017lag,Bastian2021-ga}, originally adopted from a relativistic mean-field theory with density-dependent meson-nucleon coupling constants (DD2; \citealp{TYPEL1999331,Hempel:2009mc,PhysRevC.81.015803}) with a string-flip microscopic quark-matter model~\citep{Kaltenborn2017-ch}. Repulsive higher-order quark-quark interactions give rise to additional pressure contributions with increasing densities~\citep{2015ApJ...810..134K}. 
A vector interaction potential in the quark phase supports high maximum masses for neutron stars~\citep{Kaltenborn2017-ch} and twin stars~\citep{Benic_2014}. 
The phase transition between hadronic and quark matter is modelled via the Gibbs construction (global charge neutrality in the mixed phase).

\subsubsection{STOS-B145 EoS}
As a second EoS, we use the Shen Bag model (in the following we interchangeably use the abbreviation STOS/STOS-B145) from~\href{www.compose.obspm.fr}{COMPOSE}~\citep{Shen1998-jc,SHEN1998435,Sagert2010-yw,Sagert2009-nd,SUGAHARA1994557}. 
The STOS-B145 EoS uses a relativistic mean-field approach for the hadronic phase. Quark matter is described by the thermodynamic Bag model containing  u, d, and s quarks %\citep{Bogolioubov1968-nf,Chodos1974-fc}
\citep{Farhi1984-bw,greiner_1987}. The Bag model is extended by the inclusion of first-order corrections to the strong coupling constant $\alpha_\mathrm{s}$ \citep{Sagert2009-nd,Sagert2010-yw}. The strong coupling constant $\alpha_\mathrm{s}$, bag parameter $B$, and strange quark mass together determine the critical density for the mixed phase. The phase transition region is constructed via the Gibbs condition where both phases in the phase co-existing region have globally conserved charge. We use a bag parameterizations~$B=145\,\mathrm{MeV}$ and a coupling constant~$\alpha_\mathrm{s}=0.7$. The squared speed of sound of the Bag model in the pure quark phase is $1/3\, c^2$. 
The maximum gravitational mass for cold matter in $\beta$-equilibrium is~$2.01 \text{M}_\odot$.
The corresponding $M$-$R$ curve has two maxima, which are connected by an unstable branch leading to the twin star phenomenon \citep{Alford2013-cb}. The first maximum (at larger radii) can reach up to~$M \sim 2.5 \text{M}_\odot$~\citep{Zha2021-xk}\footnote{depending on the specific entropy per Baryon in the core}.

\subsubsection{CMF EoS}
The Chiral SU(3)-flavour parity-doublet Polyakov-loop quark-hadron mean-field model (CMF) combines a mean-field description of the interaction between the lowest baryon octet (p, n, $\Lambda$, $\Xi^-$, $\Xi^0$, $\Sigma^-$, $\Sigma^0$, $\Sigma^+$), the three light quark flavours u, d, s, and gluons, as well as contributions of full Hadron-Resonance list. Both the lowest octet hadrons and quarks interactions are modelled by a chiral Lagrangian \citep{Papazoglou1999-zm, Steinheimer2011-ss,Motornenko:2019arp} which allows for chiral symmetry restoration in the hadronic sector as well as in the quark sector. In addition, the thermal contributions of all other known hadronic species, including mesons and baryonic resonances, are included to properly describe QCD matter at intermediate energy densities. The CMF model thus provides a most complete description of the interactions in both the hadronic and the deconfined phase of QCD. In the CMF model, the transition between hadronic matter and quark matter is introduced by an excluded-volume formalism. The parameters of the model are chosen such that properties of nuclear matter are reproduced and the model describes lattice-QCD thermodynamics results. The model therefore incorporates a first-order nuclear liquid-vapor phase transition at densities $\mathord{\sim} \rho_\mathrm{sat}$; a second, but weak first-order phase transition occurs, due to chiral symmetry restoration, at about $4 \rho_\mathrm{sat}$ with a critical endpoint at $T_\mathrm{CeP}\approx 15\,\mathrm{MeV}$. The transition to quark matter at higher densities occurs as a smooth crossover~\citep{Motornenko:2019arp}. At asymptotically high densities, the squared speed of sound approaches the Stefan-Boltzmann limit of $\mathord{\sim}1/3 \,c^2$. The CMF model predicts hybrid neutron stars with gravitational masses up to $\mathord{\sim} 2 \text{M}_\odot$ for cold neutron star matter in $\beta$-equilibrium. The smooth nature of the crossover from hadrons to quarks does not lead to a third family branch of compact stars. 
All these components allow the CMF model to be applied for modeling of heavy-ion collisions~\citep{Steinheimer:2009nn,OmanaKuttan:2022the}, analysis of lattice QCD data~\citep{Steinheimer_2011, Steinheimer_2014,Motornenko:2019arp,Motornenko:2020yme,Motornenko2020-gg}, as well as studies of cold neutron stars and their mergers~\citep{most_2022}.
Since the CMF EoS does not include heavy nuclei at sub-saturation density, we extend it to low densities using the SFHx EoS \citep{steiner_13}, which is matched to the CMF table at densities less than
$8\times 10^{13}\,\mathrm{g}\,\mathrm{cm}^{-3}$. Note that the CMF model with a similar matching has been recently applied to describe the dynamical evolution of binary neutron star mergers as well as heavy ion collisions at the \texttt{SIS18} accelerator \citep{most_2022}.

\subsection{Supernova Simulations: Numerical Methods}
For our supernova simulations, we employ the finite-volume neutrino hydrodynamics code \code{CoCoNuT-FMT} for solving the general-relativistic equations of hydrodynamic in spherical symmetry in Eulerian form
\citep{mueller_10,mueller_15}.
The hydrodynamics module \code{CoCoNuT} uses higher-order piecewise parabolic reconstruction \citep{1984JCoPh..54..174C}
and the relativistic HLLC Riemann solver \citep{mignone_05_a}. We treat convection in 1D using mixing-length theory as applied previously
in supernova simulations \citep{wilson_88,mueller_15b,mirizzi_16}.

For the neutrino transport we use the fast multi-group transport \code{FMT} scheme of \citet{mueller_15} which solves the energy-dependent neutrino zeroth moment equation for electron neutrinos, electron antineutrinos, and heavy-flavour neutrinos in the stationary approximation using a one-moment closure from a two-stream Boltzmann equation and an analytic closure at low optical depth. Neutrino interaction rates include absorption and scattering on nuclei and nucleons and bremsstrahlung for heavy-flavour neutrinos in a one-particle rate approximation; see
\citet{mueller_15,mueller_19a} for details.
The appearance of quarks is expected to decrease neutrino opacities in the core \citep{steiner_01,pons_01,colvero_14}, which will impact the neutrino emission and any wind outflows from the PCS on longer time scales. However, one can argue \citep{Fischer2010-hy} that over short time scales after the second bounce, neutrino trapping is still effective in the deconfined core region. As a pragmatic solution, we therefore use the nucleonic opacities throughout, assuming a nucleonic composition compatible with charge neutrality (see below).
A proper neutrino treatment for the mixed phase and pure quark phase should be further explored in future studies.

Different treatments for the EoS and nuclear reactions are applied in various regimes. At low densities, the matter is treated as a mixture of electrons, positrons, photons, and a perfect gas of nucleons and nuclei. At temperatures below $5\,\mathrm{GK}$, we employ a flashing treatment for nuclear reactions following \citep{2002A&A...396..361R}; above $5\,\mathrm{GK}$ we assume nuclear statistical equilibrium. At high densities, a tabulated nuclear EoS is used. We track the mass fractions of protons, neutrons, $\alpha$-particles, and 17 nuclear species,
and the electron fraction $Y_\mathrm{e}$ in all EoS regimes.
The transition density between the low- and high-density EoS regime is set to $5\times 10^8\,\mathrm{g}\,\mathrm{cm}^{-3}$ during the collapse phase and changed to $10^{11}\,\mathrm{g}\,\mathrm{cm}^{-3}$ after the collapse.
Since our neutrino transport presently does not use consistent opacities for the quark phase, we do not add separate advection
equations of the mass fractions of u, d, and s quarks.
This is possible since the mass fractions merely act as passive scalars in the high-density EoS regime and do not influence the solution of the equations of hydrodynamics.
Instead, we map u, d, and s quarks into neutrons and protons such as to ensure charge neutrality and baryon number conservation, i.e.,
 \begin{align*}
     \tilde{X}_\mathrm{n} &= X_\mathrm{n} + \frac{1}{3} X_\mathrm{u} + \frac{2}{3} (X_\mathrm{d}+X_\mathrm{s}), \\
     \tilde{X}_\mathrm{p} &= X_\mathrm{p} + \frac{2}{3} X_\mathrm{u} + \frac{1}{3} (X_\mathrm{d}+X_\mathrm{s}),
 \end{align*}
 where $X_\text{u}$, $X_\text{d}$, $X_\text{s}$,
 $X_\text{n}$, and $X_\text{p}$  are the mass fractions of u, d, s quarks, neutrons, and protons in the EoS table, and 
 $\tilde{X}_\text{n}$ and $\tilde{X}_\text{p}$ are the mass fractions of neutrons and protons used in the code.

\subsection{Progenitor Models}
We use 40 progenitors in the mass range $14$--$100\,\mathrm{M}_\odot$ with two different metallicities $Z=0$ and $Z=0.012$ (solar metallicity). Models that start with the letter \texttt{s} (e.g., \texttt{s14}) have solar metallicity, models that start with \texttt{z} have zero metallicity. The number in the model label denotes the ZAMS (zero-age main sequence) mass in solar masses. The progenitors have been calculated with the stellar evolution code \textsc{Kepler} \citep{weaver_78,heger_10}. The solar-metallicity models are a subset of those in \citet{mueller_16a}.

Our progenitors differ in several respects from the stellar evolution models used in~\citet{Fischer2017lag, Fischer2020-rg,Fischer2021-zh}, which were taken from~\citet{umeda_2008}. Their models were based on stellar evolution calculations using a different treatment of mixing processes (Schwarzschild criterion instead of Ledoux). Furthermore, all of the progenitors from \citet{umeda_2008} have very low but non-zero metallicity (\texttt{$Z=\mathrm{Z}_\odot/200$}).
In terms of mass loss, there is no appreciable difference to our zero-metallicity models; mass loss will be negligible in both cases.
The more relevant parameters for comparison are the He, CO, and Fe core masses, which have significant influence on the supernova dynamics and are strongly dependent on the physics treatment during stellar evolution calculations. The aforementioned differences shift the relationship of core masses and the corresponding ZAMS mass. Our progenitors show a trend towards lower He, CO, and Fe core masses for a given ZAMS mass. 
More specifically, the $50\,\mathrm{M}_\odot$ progenitor \texttt{z50} in our study has a He core mass of $17.78\,\mathrm{M}_\odot$ as opposed to $21.8\,\mathrm{M}_\odot$ for the same ZAMS mass in \citet{umeda_2008,Fischer2017lag}, a CO core mass of $11.34\,\mathrm{M}_\odot$ as opposed to
$19.3\,\mathrm{M}_\odot$, a Fe core mass of $1.86\,\mathrm{M}_\odot$ as opposed to $2.21\,\mathrm{M}_\odot$, and a core binding energy $2.10\times 10^{51}\,\mathrm{erg}$ instead of $3.67\times 10^{51}\,\mathrm{erg}$. We find that the $50\, \mathrm{M}_\odot$ model 
of \citet{umeda_2008,Fischer2017lag}
corresponds most closely to our \texttt{z60} model with a He, CO and Fe
mass of
$23.90\,\mathrm{M}_\odot$,  $20.90\,\mathrm{M}_\odot$, and $2.01\,\mathrm{M}_\odot$, respectively, 
and a binding energy of $3.42\times 10^{51}\,\mathrm{erg}$.

\subsection{Nucleosynthesis Post-Processing}

To follow the nucleosynthesis we post-process the recorded temperature, density, and radius trajectories along with the neutrino fluxes and energies for $\nu_{\mathrm{e}}$, $\bar{\nu}_{\mathrm{e}}$, and $\nu_\mathrm{x}$ neutrinos.  The $\nu_\mathrm{x}$ species stands for the sum of the $\nu_{\mu}$, $\bar{\nu}_{\mu}$, $\nu_{\tau}$, $\bar{\nu}_\tau$ contributions. 
The local neutrino energy density and mean energy at the current location of each mass shell are taken directly
from the hydrodynamics code.

After the end of the hydrodynamic simulations at time $t=t_\mathrm{h}$, we extrapolate the trajectories to time $t_\mathrm{f}=1\,\mathrm{yr}$ after the explosion assuming adiabatic homologous expansion where we crudely estimate the velocity from the current radial coordinate and the time since the onset of core collapse:
\begin{eqnarray}
T(t) = T(t_\mathrm{h}) \times \left(\frac{t}{t_\mathrm{h}}\right)^{-1}
\;,\quad
\rho(t) = \rho(t_\mathrm{h}) \times \left(\frac{t}{t_\mathrm{h}}\right)^{-3}
\;.
\end{eqnarray}
We additionally impose a minimum temperature of $10^6\,\mathrm{K}$.  At that stage, all regular nuclear reactions are frozen out, only radioactive decays still occur, which we do follow.

For the neutrinos we assume that the neutrino luminosities, $L_\nu$, and energies, $E_\nu$, decay exponentially after time $t_\mathrm{h}$, 
\begin{eqnarray}
L_\nu(t) &=& L_\nu(t_\mathrm{h}) \times \exp\left(\frac{t_\mathrm{h}-t}{\tau_\nu}\right)
\;,\quad
\\
E_\nu(t) &=& E_\nu(t_\mathrm{h}) \times \exp\left(\frac{t_\mathrm{h}-t}{4\,\tau_\nu}\right)
\;,
\end{eqnarray}
using a characteristic timescale of $\tau_{\nu}=3\,\mathrm{s}$.   The neutrino temperatures (in MeV) are approximated from the neutrino energies as $T_\nu=E_\nu / 3.15$.  The neutrino temperatures in the network are explicitly limited to a range of $3$-$8\,\mathrm{MeV}$ for $\nu_\mathrm{e}$, to $3$-$12\,\mathrm{MeV}$ for $\bar{\nu}_\mathrm{e}$, and to $4$-$12\,\mathrm{MeV}$ for $\nu_\mathrm{x}$. 

We use a modified standalone version of the first-order implicit adaptive nuclear reaction network from \textsc{Kepler} \citep{2002ApJ...576..323R}, which includes all nuclear species and reactions up to astatine except fission. The network also includes neutrino-induced spallation as described in \citep{2005PhLB..606..258H}. 
Compared to the \textsc{Kepler} version,
the standalone version (\textsc{Burn} code) of the network adds new features for higher computational accuracy.
Due to the potentially large time step in the trajectories, we implement a new \emph{iterative} adaptive network that repeats a time step until there is no more addition of new species rather than only adjusting the network at the end of a time step in preparation for the next time step.  We also iteratively sub-cycle the network calculation when the abundance changes are too large or when mass conservation is violated by more than one in $10^{-14}$, and linearly interpolate thermodynamic and neutrino quantities from the recorded or extrapolated trajectory grid points. 

For temperatures exceeding $10^{10}\,\mathrm{K}$ we impose the $Y_\mathrm{e}$ from the hydrodynamics code assuming a composition with free nucleons only.  When the temperature drops below this threshold temperature, we follow the full network using this new $Y_\mathrm{e}$ as a starting point.  This accounts for the limited temperature range for neutrino interactions in \textsc{Kepler} and takes advantage of the full non-thermal neutrino energy distribution in \textsc{CoCoNuT}. 

\section{Results}\label{section three}

\begin{figure}
    \centering
    \includegraphics[trim=5mm 5mm 2mm 1mm,
    width=\linewidth]{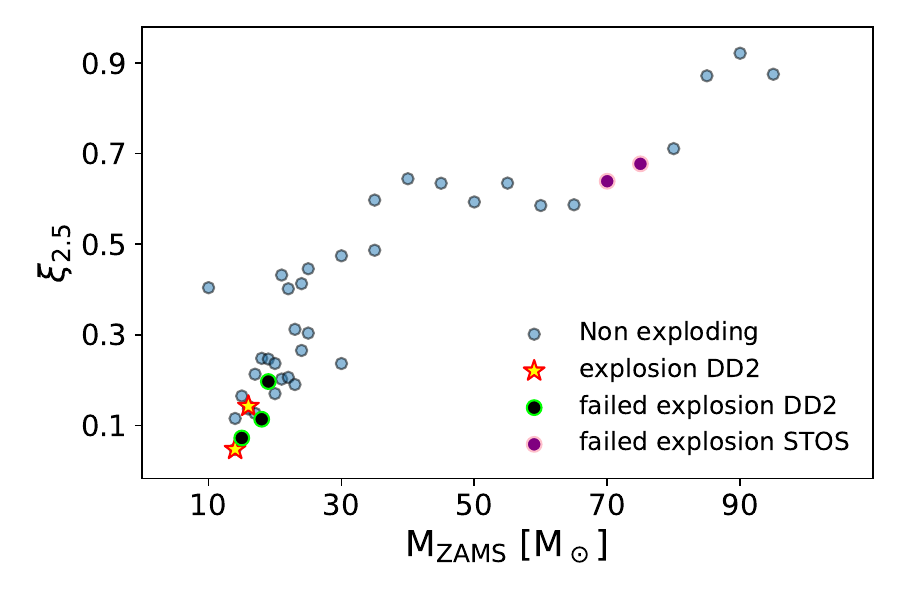}
    \caption{Compactness parameter $\xi_{2.5}$ for the set of progenitors used in our simulations. Yellow stars denote successful explosions with the DD2F\_SF EoS, black dots with green/pink edging denote failed explosions for DD2F\_SF/STOS where the shock wave is not powerful enough to propagate through the infalling material or the maximum compact star mass is exceeded due to accretion during the explosion phase. Blue circles comprise all other non-exploding progenitors.}\label{fig:comp}
\end{figure}

We summarise key outcomes from all simulations in
tables~\ref{table:dd2 2},~\ref{table:shen table}, and~\ref{table:cmf}, including
the time of the phase transition, the presence or
absence of a second bounce and neutrino burst,
the central lapse function $\alpha$ at the end of the simulation, and the occurrence or absence 
of an explosion. In addition to the zero-age  main sequence (ZAMS) mass of the models, it is useful to consider the compactness parameter $\xi_{2.5}$ at $t=0$ where
\begin{equation}
    \xi_M=\frac{M/\text{M}_\odot}{r(M)/1000\,\text{km}},
\end{equation}
for any mass coordinate $M$. The compactness parameter is one  empirical predictor for the likelihood of a star to explode by the neutrino-driven mechanism \citep{OConnor2011-le}, but it will also prove useful as a stellar structure metric in the context of PT-driven explosions. The compactness parameters
for all progenitor models are also plotted in Figure~\ref{fig:comp}. In addition, the tables show the explodability parameters $M_4$ and $\mu_4$ of \citet{ertl_16}, which are related to the mass of the iron-silicon core (using the mass shell where
the entropy $s$ reaches $4k_\mathrm{B}/\mathrm{baryon}$)
and the density outside the Si/O shell interface.

\begin{table*}
\input{tables/table_dd2_1.4}
\caption{Progenitor properties and results for the 1D simulations using the DD2-1.4 EoS. $M$ is the ZAMS mass, $M_\mathrm{PCS}$ is the mass of the PCS at the time of the second collapse, $t_\mathrm{BH}$ is the time of BH formation (models that did not form BH after $t\geq > 6\mathrm{s}$ are marked with  ``--''), $t_\text{1,b}$ is the time of the first bounce, $t_\text{MP}$ is the time when the progenitor reaches the mixed-phase region, $t_{2,\mathrm{b}}$ is the time of the second bounce, $\alpha$ is the central lapse function at the end of the simulation, $\xi_{2.5}$ is the progenitor compactness, $\mu_4$ and
$M_4 \mu_4$ are the explodability parameters from \citet{ertl_16} (evaluated at $t=0$). The column $C^\mathrm{inv}_\mathrm{L}$ refers to the presence or absence of inverted convection after the phase transition.  
}
\label{table:dd2 2}
\end{table*}

\begin{table*}
\input{tables/table_stosB145.tex}
\caption{Properties of the 1D simulations for 21 different progenitors using the STOS B145 EoS. $M$ is the ZAMS mass, $M_\mathrm{PCS}$ is the mass of the PCS at the time of the second collapse, $t_\text{1,b}$ is the time of the first bounce, $t_\text{MP}$ is the time when the progenitor reaches the mixed-phase region, $t_{2,\mathrm{b}}$ is the time of the second core bounce, $\alpha$ is the central lapse function at the end of the simulation, $\xi_{2.5}$ is the progenitor compactness, $\mu_4$ and
$M_4 \mu_4$ are the explodability parameters from \citet{ertl_16} (evaluated at $t=0$). $C^\mathrm{inv}_\mathrm{L}$ refers to the presence (+) or absence (-) of inverted convection after the phase transition.}
\label{table:shen table}
\end{table*}

\begin{table*}
\input{tables/table_cmf.tex}
\caption{Summary of models (analogous to Table~\ref{table:dd2 2}) using the CMF EoS. $M$ is the ZAMS mass, $M_\mathrm{PCS}$ is the mass of the PCS at the time of the second collapse, $t_\mathrm{BH}$ is the final time of the simulation ($t_\mathrm{BH}$ is the time of BH formation (models which cooled before reaching the crossover region and did not collapse to a BH within $t \leq 6\mathrm{s}$ are marked with ``-''), $t_\text{1,b}$ is the time of the first bounce, $t_{2,\mathrm{b}}$ is the time of the second core bounce, $\alpha$ is the central lapse function at the end of the simulation, $\xi_{2.5}$ is the progenitor compactness, $\mu_4$ and
$M_4 \mu_4$ are the explodability parameters from \citet{ertl_16} (evaluated at $t=0$). $C^\mathrm{inv}_\mathrm{L}$ refers to the presence (+) or absence (-) of inverted convection after the phase transition. Some progenitors show very slight convection (with a negative entropy gradient).}
\label{table:cmf}
\end{table*}

Among all 97 models, we find only two explosions by the PT-driven mechanism, both for the DD2F\_SF EoS, namely the zero-metallicity model \texttt{z14} and the solar-metallicity model \texttt{s16}. We, therefore, consider the DD2F\_SF series first, with a particular focus on the two exploding models.

\subsection{DD2F\_SF Series}\label{subsection:DD2}
\subsubsection{Progenitor-Dependent Outcomes}
The two exploding models \texttt{z14} and \texttt{s16} point to a remarkable difference from the scenario of hyperenergetic explosions from very massive progenitors in \citet{Fischer2017lag}.
Those two models both have low compactness parameters $\xi_{2.5} < 0.15$ (marked as yellow stars in Figure~\ref{fig:comp}).
Furthermore, they only reach low explosion energies.
\begin{figure}
    \centering
    \includegraphics[trim=5mm 5mm 5mm 5mm,
    width=0.8\columnwidth]{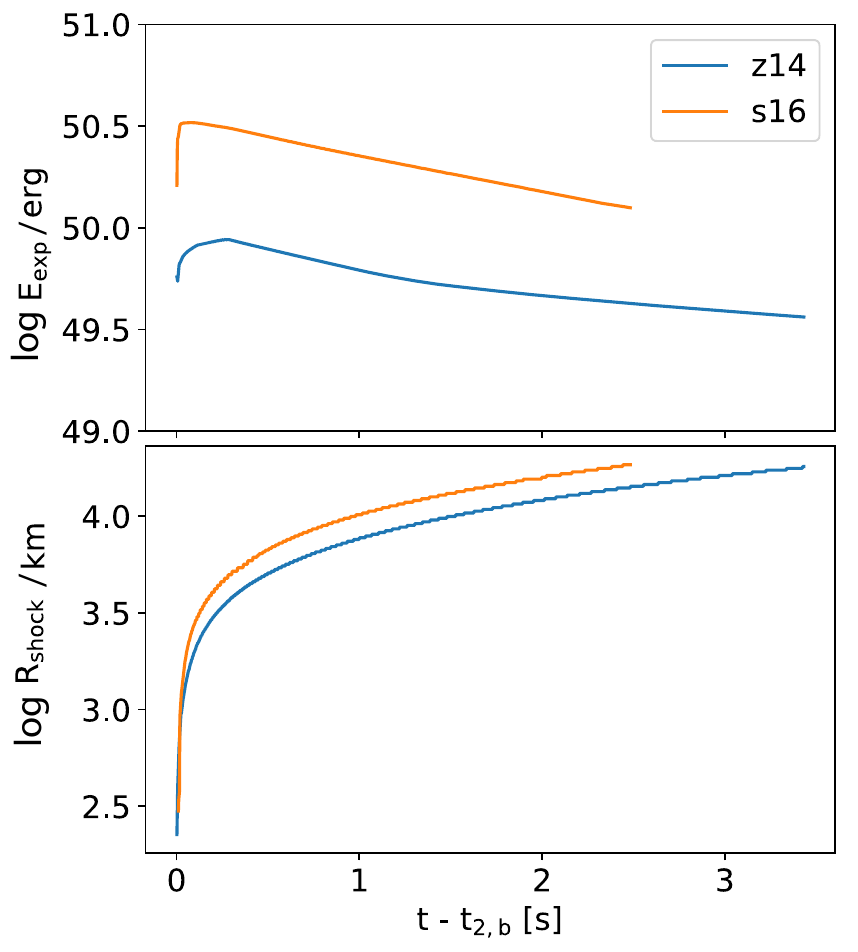} 
    \caption{Diagnostic explosion energy $E_\mathrm{expl}$ (top) and shock radius $R_\mathrm{shock}$ (bottom) as function of time after the second bounce for models \texttt{z14} (orange) and \texttt{s16} (blue) using the DD2F\_SF EoS. By the end of the simulation, we find diagnostics energies     of $1.25\times 10^{50}\,\mathrm{erg}$ and $3.64\times10^{49}\,\mathrm{erg}$ for
    \texttt{s16} and \texttt{z14}, respectively.
    In both cases, much of the initial explosion energy is drained as the shock scoops up bound shells. 
    \label{fig:expl}}
\end{figure}
In Figure~\ref{fig:expl} we show the evolution of the diagnostic explosion energy (computed in general relativity following \citealt{mueller_12a}) and shock trajectories for these two models.
We find values of only
$1.3\times~10^{50}\,\mathrm{erg}$ for \texttt{z14} and $3.6\times~10^{49}\,\mathrm{erg}$ for \texttt{s16} at the end of the simulations. Model \texttt{s16}, transiently reaches a diagnostics energy of about $3\times 10^{50}\, \mathrm{erg}$ immediately after the second bounce, but the diagnostic energy then steadily decreases as the shock scoops up bound outer layers of the star. In model \texttt{z14}, the shock launched by the second bounce is noticeably weaker; the diagnostic energy never exceeds $10^{50}\, \mathrm{erg}.$

Not all of the other models form black holes quietly, however.
Three models in the DD2F\_SF series turn out as \emph{failed explosions}.
We call an explosion failed if there is a second bounce
after the phase transition and the second shock initially propagates
dynamically with positive post-shock velocities through the neutrinosphere that leads to a second neutrino burst, but then proves too weak to propagate through outer infalling material and stalls again. 
Similar to \citet{Zha2021-xk} we note oscillating behaviour for the failed explosion model \texttt{z18}
in which the PCS oscillates for several milliseconds before it collapses into a BH.
The trend towards low compactness parameters for failed explosions is similar to exploding models- we only see failed explosions for $\xi_{2.5} \lesssim 0.2$ (models \texttt{z15}, \texttt{z18}, and \texttt{z19}). 

In order to gain more insight into the influence of the phase transition after the second collapse,  we consider the pressure-weighted mean adiabatic index ${\Gamma}$ as a measure for the
stability of the PCS \citep{Goldreich1980-xm},
\begin{equation}\label{gamma}
    \overline{\Gamma} = \frac{\int_\text{PCS} \Gamma\, P\,\text{d}V}{\int_\text{PCS} P\,\text{d}V},
\end{equation} 
where the integration volume covers the PCS, which we define by a 
threshold density of $10^{11}\,\mathrm{g}\, \mathrm{cm}^{-3}$. The relation between adiabatic index and stellar stability is well known~\citep{Chandrasekhar1964-gs}. For spherical stars in non-isotropic matter, instability can be shown to set in when the pressure-weighted adiabatic index decreases below $4/3$~\citep{Goldreich1980-xm}, though corrections
apply in general relativity. The adiabatic index usually decreases during phase transitions. It is apparent that this softening causes the initial contraction of the PCS. 

\begin{figure}
    \centering
    \includegraphics[width=\linewidth]{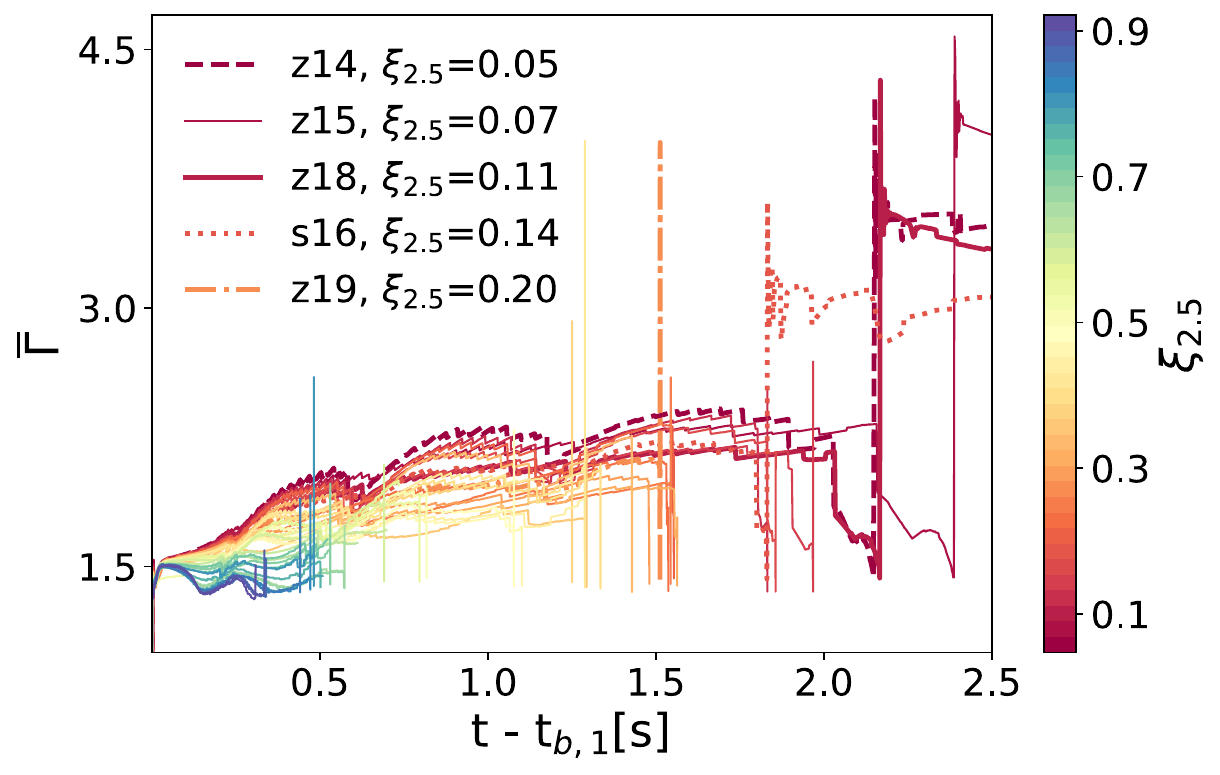}
    \caption{Evolution of the mean adiabatic index $\bar{\Gamma}$ (defined in Equation \ref{gamma}) as function of post-bounce time, colour-coded by the compactness parameter, for the DD2F\_SF models. The thick (dashed or dotted) lines show progenitors that either exploded (\texttt{z14}, \texttt{s16}), or failed to explode (\texttt{z15}, \texttt{z18}, \texttt{z19}) after the emission of a second neutrino burst due to the stalling of the second shock .\label{fig:four}}
\end{figure}
We plot $\bar{\Gamma}$ as function of  time after the first bounce in Figure~\ref{fig:four}. The progenitors are colour-coded by their compactness parameter~$\xi_{2.5}$ at~$t=0$.
High-compactness progenitors reach the phase transition earlier due to the shorter free-fall time scale of the shells outside the iron core.
High-compactness progenitors also show a lower mean adiabatic index compared to low-compactness progenitors. This makes high-compactness progenitors more susceptible to bulk collapse to a black hole (without a second bounce) as soon as they hit the phase transition. However, for the DD2F\_SF EoS, only 3 out of 40 progenitors do \emph{not} exhibit a second bounce.
The two exploding models \texttt{s16} and \texttt{z14} (thick dashed and dotted curves, respectively) exhibit some of the highest values of $\bar\Gamma$, although not the highest pones altogether.
As a reference, we also mark the failed CCSNe as thick dashed/dotted lines. The dependence of $\bar\Gamma$ on the compactness parameter reflects the trend towards higher average PCS entropy in high-compactness parameters noted by \citet{Schneider_2020,Zha2021-xk}. 
In addition, the higher chance of a PT-driven explosion for low-compactness progenitors is also consistent with another phenomenon described by \citet{Zha2021-xk}, who noted that
the phase transition for oscillating progenitors occurs well below the maximum mass for a stable PCS, and at lower average PCS entropy. At the time of the second collapse, the PCS masses of the exploding models are $1.71 \text{M}_\odot$ for model \texttt{z14} and $1.76 \text{M}_\odot$ for model \texttt{s16}.

\begin{figure*}
	\centering
	\includegraphics[trim=5mm 5mm 5mm 5mm,
	width=\linewidth]{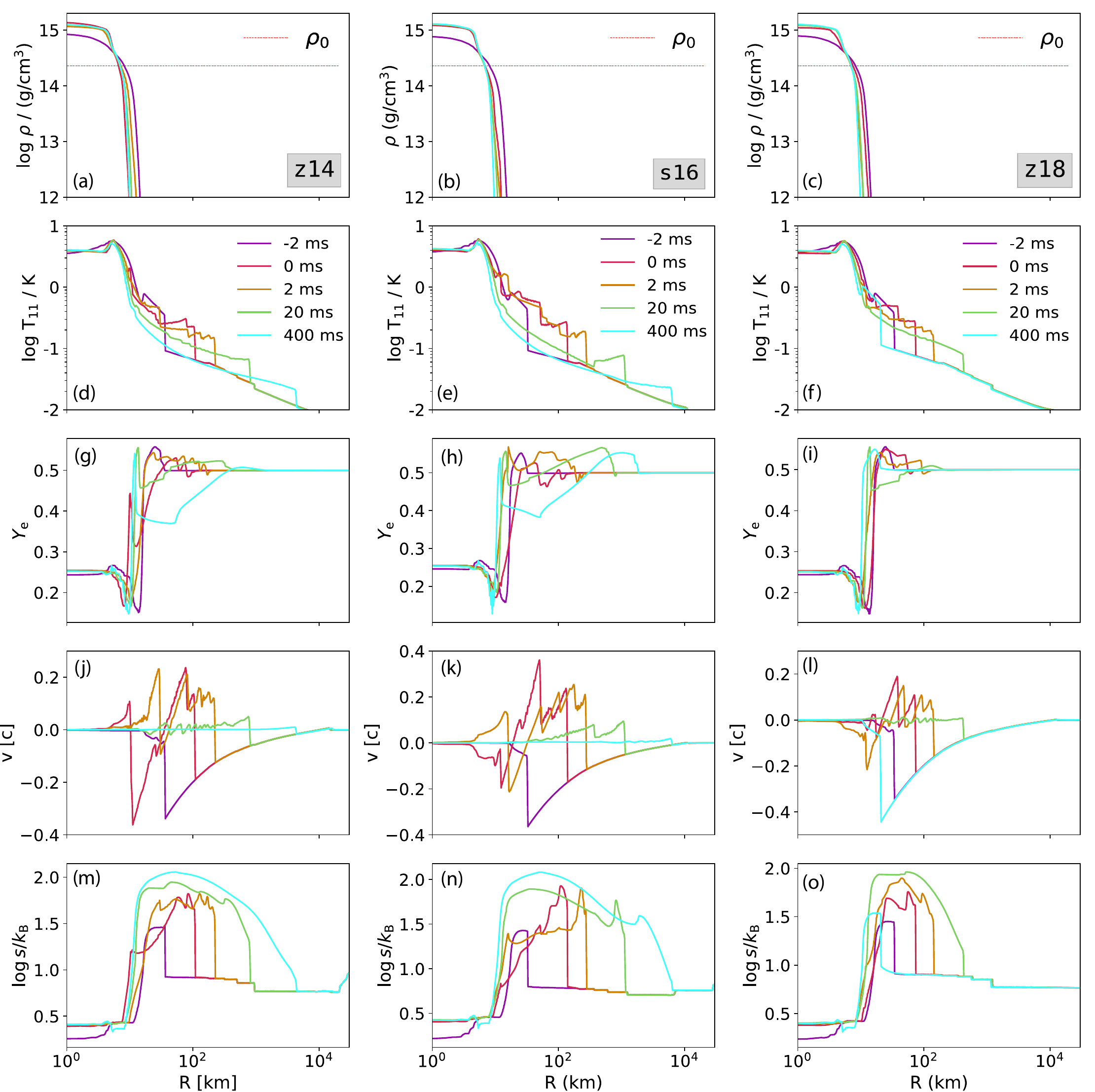}
	\caption{Selected radial profiles for \texttt{z14}, \texttt{s16}, and \texttt{z18} around the time of second collapse using DD2F\_SF. The panels show (from top to bottom) density $\rho$, temperature $T$
	in $10^{11}\,\mathrm{K}$, $Y_\mathrm{e}$, radial velocity $v$, 
	and baryon-specific entropy $s$ as functions of radius at five different time steps with $t=0$ corresponding to when the star enters the mixed phase, which marks the onset of the second collapse.\label{fig:radial profiles}}
\end{figure*}

\subsubsection{DD2F\_SF Series -- Detailed Dynamics of Selected Models}\label{sec:dd2_hydro}
We continue our discussion by outlining dynamic properties of the exploding models \texttt{z14} and \texttt{s16} and the failed explosion model \texttt{z18}. 
Radial profiles of the density $\rho$, temperature $T$, electron fraction $Y_\mathrm{e}$, radial velocity $v$, and entropy per baryon $s$\footnote{In the mixed phase and quark phase, ``entropy per baryon'' is to be understood as the entropy of a fluid element divided by its net baryon number, which can still be defined even when quarks are not bound in baryons anymore.}
at five epochs
 from $2\,\mathrm{ms}$ before the star reaches the mixed phase up to 400~ms
after the second bounce are shown in Figure~\ref{fig:radial profiles}.

Shortly before the centre of the PCS enters the mixed phase (purple curves), continuous accretion has brought the central density up to several times saturation density, and the density, temperature, and entropy exhibit typical steady-state accretion profiles. Strong exposure to neutrino heating raises $Y_\mathrm{e}$ in the accreted matter to values above $0.5$ in much of the gain region, before it drops as material settles in the cooling region. This will later become relevant for nucleosynthesis.
Once the mixed phase is reached, the evolution becomes extremely fast. On timescales of less than a millisecond, the PCS goes through rapid, near-homologous contraction due to the lower adiabatic index in the mixed phase, and due to increasing densities, a major part of the PCS is soon composed of pure quark matter. By the time corresponding to the red curves, the collapsing PCS has already undergone a rebound due to increasing quark-vector interactions, halting the collapse at about $\sim 2\times 10^{15}\,\,\mathrm{g}\,\mathrm{cm}^{-3}$. 
The vector interactions in the quark phase go along with violations of lattice QCD data which we briefly touched upon in Section~\ref{section two}. 
The shock wave from the second bounce has already broken out of the PCS, overtaken the primary shock, and reached a radius of order $100\, \mathrm{km}$, heating the post-shock matter to $40\texttt{-}80 \, k_\mathrm{B}/\mathrm{baryon}$. The velocity profiles around the time of the second bounce (red, dark yellow) still show strong ringdown oscillations of the PCS, which launch further secondary shocks that somewhat boost the entropy in the ejected matter. 
Several shock waves from the bounce and ringdown oscillations are formed within the first $2\mathrm{ms}$: The shock wave of model \texttt{z14} (red line in panel j) at about $200\,\mathrm{km}$ comes from the initial \underline{second} collapse and  bounce. Another rebound occurs at $3\,\mathrm{km}$ after $\mathord{\sim} 0.5\,\mathrm{ms}$ (on shorter timescales than displayed here). Multiple core oscillations follow within time scales $\mathord{\lesssim} 2\,\mathrm{ms}$ and lead to the ``jagged'' velocity profile which we observe for successful as well as failed explosions. 

The promptly ejected matter previously located in the gain region mostly maintains $Y_\mathrm{e}>0.5$ as it is run over by the shock from the second bounce, but the profiles show variations in $Y_\mathrm{e}$ in the promptly ejected matter; this can be understood as the result of neutrino irradiation by the intense electron antineutrino burst \citep{Sagert2009-nd} from the breakout of the second shock through the neutrinosphere.

Already right after the second bounce, the failed explosion model \texttt{z18} exhibits noticeably smaller positive velocities immediately behind the  main (outermost) shock, and smaller amplitudes of the additional shock waves launched by ringdown oscillations. In this model, the shock still propagates out to several hundred kilometres  and then stalls again (panel l, green curve). The shock then recedes to a few tens of kilometres, and the PCS continues to accrete for several hundred milliseconds after the phase transition. In models \texttt{z14} and \texttt{s16}, the shock continues to propagate outwards, but the post-shock velocity decreases considerably as the initial kinetic energy of the shock is used up to unbind the shells around the PCS. Both of these models develop a neutrino-driven wind from the PCS with high entropy $\mathord{\gtrsim}100 \,k_\mathrm{B}/\mathrm{baryon}$ (panels m and n, cyan curves) in the wake of the explosion. The outflow velocities in the wind remain modest, but the developing wind can clearly be seen for model \texttt{s16} (panel k, green curve), where the wind crashes into the earlier ejecta
in a reverse shock at a few hundred kilometres. We will touch upon the dynamics of the wind and the reverse shock later in Section~\ref{sec:neutrino}, as it leaves interesting traces in the neutrino signal.

\begin{figure}
	\centering
	\includegraphics[width=\columnwidth]{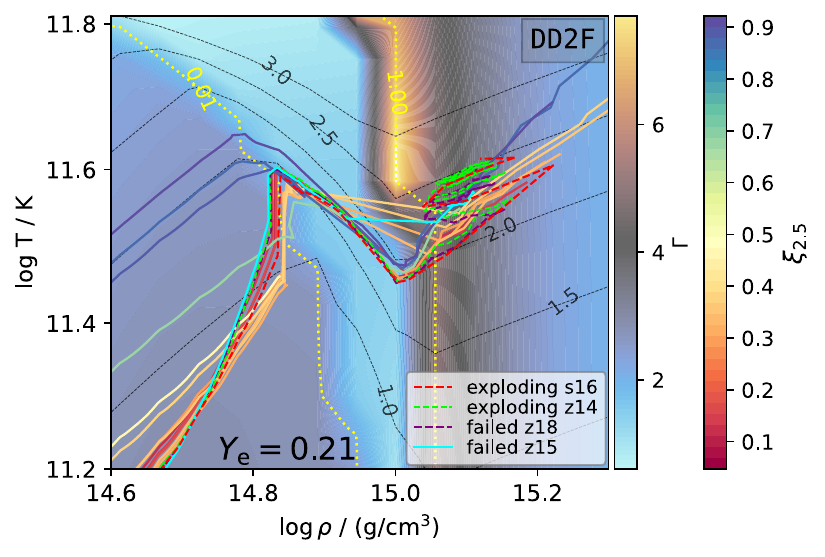}
	\caption{Phase diagram with trajectories of the central density and temperature (thick solid and dashed lines) of selected DD2F\_SF models. The two exploding progenitors are thick red and green dashed and the failed explosions of \texttt{z18} and \texttt{z15} are shown in thick dashed purple and blue, respectively. The background displays the colour-coded adiabatic index at fixed electron fraction $Y_\mathrm{e}=0.25$
	for the DD2-RMF-1.4 EoS . Two isocontours (dotted, yellow) show quark fractions of $\text{X}_\text{q}=0.01$ and $\text{X}_\text{q}=1$. The black dashed lines are isentropes for different entropy values (indicated as contour labels).  Most of the progenitors used in this setup with the exception of two models (\texttt{z20}, \texttt{z21}) that underwent a second core bounce. \label{fig:dd2 phase}. }
\end{figure}

With only two exploding models and a few more failed explosions, it is difficult to ascertain the reason for variations in the initial energy of the second shock (and hence the outcome of the phase transition).
The trajectories of the central density and temperature in the phase diagram provide some tentative hints, though. All except two progenitors in the DD2F\_SF setup undergo a second core bounce. 
The trajectories of central temperature and density of selected models from the DD2F\_SF series are shown in Figure~\ref{fig:dd2 phase} together with the
adiabatic index (background colour), adiabats, and isocontors for the quark fraction at a fixed electron fraction $Y_\mathrm{e}=0.25$, which approximates the conditions at the centre of the PCS.

After the first bounce and before the PCS enters the mixed phase, the central density and temperature of the PCS increase due to mass accretion to $\log \rho/(\mathrm{g}\,\mathrm{cm}^{-3})\approx 14.9$,~$\log (T /\mathrm{K}) \approx 11.45$. At the onset of the mixed phase, lower and intermediate progenitors show a sudden jump in temperature which is due to inverse convection as we will discuss in Section~\ref{sec:pt_eos}. 
When the PCS enters the mixed phase and the combined mass fraction of  u and d-quarks reaches $\mathord{\sim} 1\%$ (dotted yellow curve\footnote{Note that the electron fraction in the phase diagram background is fixed to $Y_\text{e} = 0.25$, so the onset of the mixed phase cannot be pinpointed exactly using the isocontours of the quark fraction.}), the trajectories bend abruptly  at approximately constant density~$\log\rho /(\mathrm{g}\,\mathrm{cm}^{-3})=14.8$. The progenitors are not yet collapsing at this point but the adiabatic index in the mixed phase decreases, softening of the EoS leads to a core contraction of the progenitors within a few milliseconds, and collapse follows once enough matter in the PCS is converted into mixed-phase matter. The adiabatic collapse is accompanied by a decrease in temperature. Such a decrease in temperature under adiabatic compression 
during a phase transition may seem peculiar, but for the QCD phase transition this behaviour can be connected to the higher entropy in the quark phase under isothermically compression (i.e., along horizontal lines in Figure~\ref{fig:dd2 phase}) as we shall discuss later in Section~\ref{sec:pt_eos}.
When the quark fraction approaches unity, the repulsive quark interactions come into play and significantly stiffen the EoS at densities~$15~\leq~\log\rho /(\mathrm{g}\,\mathrm{cm}^{-3}) \leq~15.1$.
The adiabatic index~$\Gamma$ is highest immediately after the conversion to quark matter. The models with a second bounce overshoot the 
``ridge'' of maximum $\Gamma$ to various degrees before the central density reaches its maximum -- corresponding to the bounce -- and then oscillate back and forth. The second bounce occurs at densities of~$\log\rho /(\mathrm{g}\,\mathrm{cm}^{-3}) =15.1$ for \texttt{z14} and \texttt{z18} and slightly later at~$\log\rho/(\mathrm{g}\,\mathrm{cm}^{-3})=15.22$ for \texttt{s16}.
The latter shows the largest increase in density during the collapse, implying that more gravitational binding energy is released followed by a relatively strong core bounce. This causes a larger explosion energy $E_\text{exp}=1.25\times 10^{50}\,\mathrm{erg}$, compared to \texttt{z14} with $E_\text{exp}=3.64 \times 10^{49}\,\mathrm{erg}$.
The failed explosion models reach a lower maximum density
during the second bounce. This provides a possible explanation for why explodability by the phase-transition mechanism is not related to compactness (or any other obvious structure parameter) in a simple, monotonic fashion: For a successful explosion, the second collapse must neither proceed too violently to high densities since this would result in black hole formation, but must concurrently reach sufficiently high densities to launch a strong shock wave. 
In summary, there is no straightforward way to quantitatively predict the maximum density during the second bounce and the energy of the shock wave solely from progenitor or PCS parameters.
Furthermore, CCSNe were explained as a transition from a second to a third family of hot compact stars~\citep{Hempel2016-jf}. When the gravitational mass of the PCS exceeds the maximum supported (entropy dependent) maximum PCS mass, the star collapses~\citep{Schneider_2020}. In addition, it was discussed that the mass ratio of both maxima in the mass-radius diagram can determine whether the star collapses into a BH immediately or bounces~\citep{Zha2021-xk}. This third family topology in the mass-radius diagram, is entropy dependent\footnote{As a consequence of unusual thermodynamic properties which we will discuss in more depth in Section~\ref{sec:pt_eos}}. The STOS-B145 EoS evinces this characteristic of a third family of compact stars at higher entropies.

Whereas we are not able to unambiguously pin down the exact mechanism leading to a successful explosion due to the the small number of exploding models, we emphasise that it is the particular --- although otherwise unremarkable --- core structure of the specific models that lead to the explosion.  The core structure strongly varies with mass, as pointed out by \citet{mueller_16a,Sukhbold2018,Sukhbold2014}.  Metallicity only plays a minor role, largely leading to some shift of the mass range where a core structure of similar explodability may be incurred.  Other parameters, such as stellar rotation, or modelling parameters such as uncertainties in mass loss, nuclear reaction rates, and mixing processes, may have an even stronger effect.  From stellar modelling experience, within reasonable limits, we should expect that all of these predominately shift the mass ranges in similar ways to metallicity.  Interestingly, both explosions occur in the intermediate CCSN mass range, for quite similar core structure.  The larger explosion energy 
for \texttt{s16 }is not an effect of metallicity per se. 

\begin{table*}
\begin{tabular}{lllllll}
\toprule
     & Nature of Phase Transition                         & $\log\rho^\text{onset}_\text{PT}$ (g\, cm$^{-3}$)& $\Delta\log\rho_\text{PT}$ (g\,cm$^{-3}$)& Enthalpic/entropic & $\Gamma^\mathrm{PT}_\text{min}$ & $\Gamma_\text{max}(\rho > \rho_\text{PT})$ \\
     \midrule
DD2F\_SF  & 1st order (Maxwell constr)            & 14.6-15.0        & $\sim$0.1-0.5    & yes/no             & $\sim 0.3$                     & $\sim 5.2$                     \\
STOS & 1st order (Gibbs constr)              & 14.5-14.1          & $\sim$1.1        & yes/yes            & $\sim 1.2$                      & $\sim 2.8$                     \\
CMF  & Smooth crossover (``$\infty$''-order) & 14.9-15.0         & $\sim$0.05       & yes/no             & $\sim 0.5$                     & $\sim 1.2$  \\      \bottomrule            
\end{tabular}
\caption{Properties of the mixed phases/crossover region of the DD2, STOS, and CMF EoS at fixed electron fraction $Y_\mathrm{e}=0.25$. 
The STOS EoS shows the lowest onset density $\rho^\text{onset}_\text{PT}$ of the phase transition, followed by DD2F\_SF. The CMF model has the smallest ``width'', i.e., density range
$\Delta \log \rho^\text{onset}_\text{PT}$
of the crossover region. STOS features a wide mixed-phase band, about twenty times the width of CMF crossover region. The minimum value $\Gamma_\mathrm{min}^\mathrm{PT}$
of the adiabatic index for DD2F\_SF is $\mathord{\sim}0.3$, about four times smaller than for STOS. DD2F\_SF also reaches significantly higher maximum values
$\Gamma_\mathrm{max}(\rho>\rho_\mathrm{PT})$
in the mixed phase/pure quark phase compared to STOS and CMF.}
\label{table:nucprop}
\end{table*}

\subsection{EoS comparison}
In addition to the DD2F\_SF models, we simulated the collapse of 36 progenitors using the CMF EoS and 21 progenitors using the STOS EoS. None of these models developed at PT-driven explosion. We find a second core bounce followed by a second neutrino burst for the STOS EoS for two high-compactness models \texttt{z70} and \texttt{z75}.  None of the CMF models exhibited a second bounce. In the following, we analyse the thermodynamic conditions at the centre of the PCS in these models and identify the reason for the disparate outcomes for the various EoS.
Important outcomes and parameters for the simulations using the STOS EoS
and CMF EoS are given in Tables~\ref{table:shen table} and \ref{table:cmf}. 
A comparison of the transition density $\rho_\text{PT}$ for the different EoS is
shown in Table~\ref{table:nucprop}, along with other characteristic properties
of the phase transition, such as the minimum adiabatic index.

\begin{figure}
	\centering
	\includegraphics[trim=2mm 2mm 6mm 2mm,
	clip=true,width=\columnwidth]{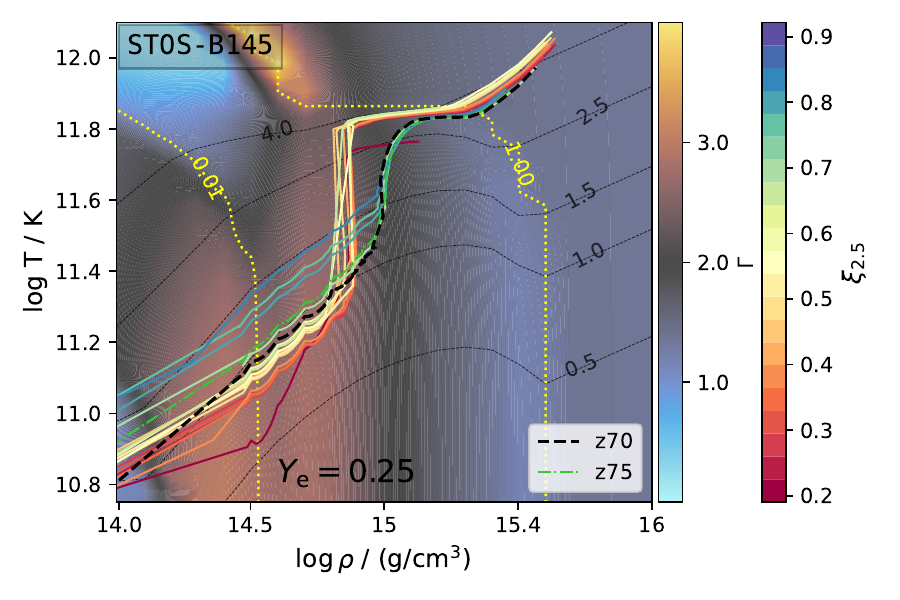}
	\caption{
	Phase diagram with trajectories of the central density and temperature (thick green and black dashed lines) of two selected STOS-Bag145 models. 
	Out of all simulations in the STOS setup, two progenitors (\texttt{z70}, \texttt{z75}) underwent a second core bounce, followed by a second neutrino burst. Those two progenitors are marked as dashed black (\texttt{z70}) and dash-dotted green (\texttt{z75})). 
	The background displays the colour-coded adiabatic index at fixed electron fraction $Y_\mathrm{e}=0.25$
	for the STOS-Bag145 EoS . Two isocontours (dotted, yellow) show quark fractions of $\text{X}_\text{q}=0.01$ and $\text{X}_\text{q}=1$. The black dashed lines are isentropes for different entropy values (indicated as contour labels). \label{fig:stos phase}}
\end{figure}
\subsubsection{STOS Models}
Similar to the DD2F\_SF phase diagram in Figure~\ref{fig:dd2 phase},
Figure~\ref{fig:stos phase} shows the evolution of the central temperature and density on top of the adiabatic index (colour-code background) and isocontours for the quark fraction at fixed electron fraction $Y_{\mathrm{e}}=0.25$.  The  appearance of quarks is shown in dotted yellow for $X_\mathrm{q} = \sum_{i\in \text{u,d,s}} X_i = 0.01$. $X_\mathrm{q}=1.0$ marks the end of the mixed phase. 
The STOS EoS features an early onset of the mixed phase at densities
$\rho_\text{PT}$ with
$14 \lesssim \log (\rho_\text{PT} /\mathrm{g}\,\mathrm{cm}^{-3}) \lesssim 14.5$ for $11.8 \lesssim \log (T/\mathrm{K})\lesssim12$. 

The transition density shifts to lower densities for higher temperatures. This trend is seen in CMF and DD2F\_SF as well and is a generic feature of the QCD phase diagram.

The adiabatic index for STOS looks qualitatively different from the DD2F\_SF EoS. For STOS the adiabatic index is roughly constant at the onset of the mixed phase with values $\Gamma\approx 3$ until $\log \rho/(\mathrm{g}\,\mathrm{cm}^{-3})~\approx~14.75$ and temperatures below $\log T/\mathrm{K} \lesssim 11.5$. By contrast, the DD2F\_SF EoS shows immediate softening at the onset of the mixed phase. The adiabatic index for the STOS EoS softens for higher densities, but does not reach values below $\Gamma=1$ as is the case for DD2. $\Gamma$ monotonically decreases with increasing density during the mixed phase and converges towards $1.2 \lesssim \Gamma\lesssim 2$ in the pure quark phase. This is significantly lower than for DD2F\_SF where $\Gamma\gtrapprox 2$. 

Isentropes show an increase in temperature at the early onset of the mixed phase and then a decrease at higher densities $\log\rho/ (\text{g\,cm}^{-3}) \gtrsim 15.4$ in the mixed phase. In contrast, DD2F\_SF featured only the latter, i.e., a decrease in temperature at constant lines of entropy.
The STOS EoS exhibits similar behaviour as DD2F\_SF only at temperatures above $\log(T/\mathrm{K}) \gtrsim 11.8$ and entropies $s \gtrsim 5
k_\mathrm{B}/\mathrm{baryon}$: The transition density to the mixed phase transitions to lower densities $14 \lesssim \log \rho_\text{PT}/ (\text{g\,cm}^{-3}) \lesssim 14.5$ and the adiabatic index significantly softens. In this regime of the mixed phase, the EoS stiffens when quarks become the dominant degrees of freedom.

The gentle softening and the lack of an abrupt increase of $\Gamma$ along
the trajectories of central density and temperature are not conducive to a strong rebound after the initiation of the second collapse in the case
of the STOS EoS.
The \emph{soft-stiff} transition due to vector repulsion in the DD2F\_SF EoS seems to play a crucial role in the dynamics of the second collapse and (where applicable) the launching of an explosion. It would be interesting to consider PCSs with high initial entropy that would cross the region of low $\Gamma$ at high temperature in the mixed phase to determine whether PT-driven explosions can occur in this regime.\footnote{Note however, that the \textit{two-EoS} description in close vicinity to the phase construction, that is where both EoS intersect, lacks reliability, since either quark or hadronic models lie beyond their regime of validity~\citep{Baym_2018}. Furthermore the \textit{two-EoS} approach does not admit a continuous phase transformation and therefore no QCD-critical end-point in the QCD phase diagram~\citep{Hempel2013-su}}, for greater detail see \S 83 in~\citet{Landau1980-ea}.

The violence of the collapse and rebound in the STOS model may also be reduced by the fact that
the thermodynamic conditions at the centre actually do not follow the adiabats. The central entropy actually increases substantially during the phase transition. 
The lower and intermediate compactness STOS models $\xi_{2.5}\lesssim 0.7$ (red/orange/yellow/light green curves in Figure~\ref{fig:stos phase}) jump in temperature during the mixed phase at a constant density of about $\log\rho/(\mathrm{g}\,\mathrm{cm}^{-3}) \approx 14.8$. Entropy values accordingly increase from $s=0.9 k_\mathrm{B}/\mathrm{baryon}$ to $3 k_\mathrm{B}/\mathrm{baryon}$.  
In the higher compactness models (blue), which start with higher central entropy, the increase in temperature and entropy is less pronounced, and the post-collapse entropy is almost progenitor-independent. This phenomenon, which may appear puzzling at first glance, is due to convective mixing as we shall discuss in Section~\ref{sec:pt_eos}.

None of the low-compactness models experience a second bounce and contract rather slowly during the mixed phase. A fast collapse follows once the progenitors pass the mixed phase region and contain a pure quark PCS core (see right thick dotted yellow line). The time between core-contraction and BH formation is roughly in between $5\, \mathrm{ms}\lesssim t_\mathrm{collapse}  \lesssim 200\, \mathrm{ms}$. 
 
The low-compactness model \texttt{z23} (red) forms an exception and instead collapses to a BH during the mixed phase with a mixed PCS core containing hadrons and quarks.

The evolution of intermediate and more massive progenitors (green/blue) with $\xi_{2.5}\gtrsim 0.7$ shows a smoother central temperature evolution lacking the sudden jump in $T$. Furthermore, their collapse occurs faster with $t_\mathrm{c}\sim 1\,\mathrm{ms}$ and at slightly lower densities \underline{during} the mixed phase. However, similar to the lower compactness models, BH formation in those models occurs as well beyond the mixed phase and with a pure quark core.
Two models in this category, \texttt{z70} and \texttt{z75} (thick dashed line), experience a second core bounce and a second neutrino burst. 
\begin{figure}
	\centering
	\includegraphics[width=\columnwidth]{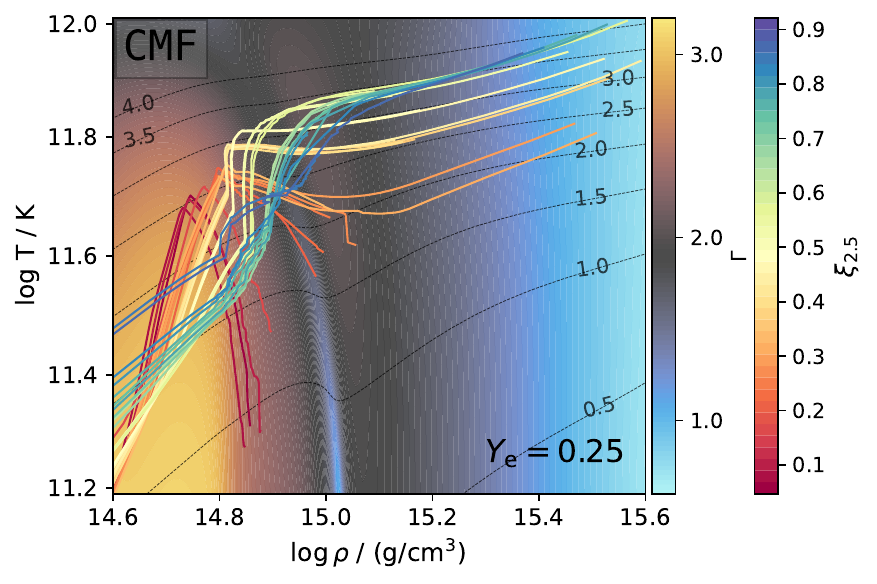}
	\caption{
		Phase diagram with trajectories of the central density and temperature (thick solid lines) of selected CMF models. 
	The background displays the colour-coded adiabatic index at fixed electron fraction $Y_\mathrm{e}=0.25$
	for the CMF EoS. The black dashed lines are isentropes for different entropy values (indicated as contour labels).\label{fig:cmf phase}}
\end{figure}

\subsubsection{CMF EoS}

We did not obtain any explosions using the CMF EoS,
and none of the models exhibits a second core bounce. Again the trajectories of central density and temperature in the phase diagram
(Figure~\ref{fig:cmf phase}) provide clues for this behaviour: 
The adiabatic index in the CMF phase diagram in Figure~\ref{fig:cmf phase} shows a smooth crossover to quarks. The onset density 
of the mixed phase (appearance of quarks)
at $\log\rho /(\mathrm{g}\,\mathrm{cm}^{-3}) \approx 15$ at fixed electron fraction $Y_\mathrm{e}=0.25$ is higher than for DD2F\_SF ($\log\rho_\mathrm{onset\,MP} \approx 14.95$ log g cm$^{-3}$) and STOS ($\log\rho_\mathrm{onset\,MP} \approx 14.5$ log g cm$^{-3}$).
Similar to DD2F\_SF and STOS, the transition density to quarks shifts to lower density at higher temperatures. The crossover region shows a decrease in temperature with increasing density along isentropes and the change in adiabatic index becomes less pronounced at higher temperatures, which is also an effect of the appearance of mesons in the hadronic and gluons in the deconfined phase. Along the actual trajectories, the adiabatic index decreases gently like for the STOS EoS without substantial stiffening after the phase transition.
The width of the crossover region at $\log (T/\mathrm{K}) \lesssim 11.8$ is significantly shorter than the mixed phases in DD2F\_SF and STOS.
The evolution of the progenitors shows four distinct types: 1) Low-compactness progenitors $\xi_{2.5} \lesssim 0.3$ (red) start to cool after $t>6\,\mathrm{s}$ and before they reach the onset of the crossover region, and simply do not undergo a second collapse.
The CMF EoS leads to a PCS structure that generally requires a larger PCS mass to reach the phase transition, especially for low-compactness models (see third column for $M_\mathrm{PCS}$ in
Tables~\ref{table:dd2 2} and \ref{table:cmf}). Therefore the time to the phase transition is generally longer for the CMF EoS than
for DD2F\_SF and STOS, which allows neutrino cooling of the PCS to become relevant before the phase transition is reached.
2) Progenitors with slightly higher compactness parameters $0.3 \lesssim \xi_{2.5} \lesssim 0.4$ (orange) show a small decrease in central temperature along isentropes during the crossover region, i.e., they cool while they move smoothly through the phase transition. 3) The intermediate- and high-compactness models with $\xi_{2.5}\gtrsim 0.5$ resemble the evolution of progenitors using the STOS EoS. The Type 3) models are heated at the centre during the phase-transition by inverted convection like the STOS models.
Type 3a) $\xi_{2.5} \gtrsim 0.5$ exhibit a jump in temperature while 3b) show a smoother temperature increase. The collapse time of 3a) and 3b) is similar, contrary to the STOS models where high compactness models with the shape of 3b) collapsed significantly faster and at lower densities. 

\subsubsection{Inverted convection and character of the phase transition}
\label{sec:pt_eos}
\begin{figure}
	\centering
	\includegraphics[trim=10mm 3mm 10mm 5mm, width=0.9\linewidth]{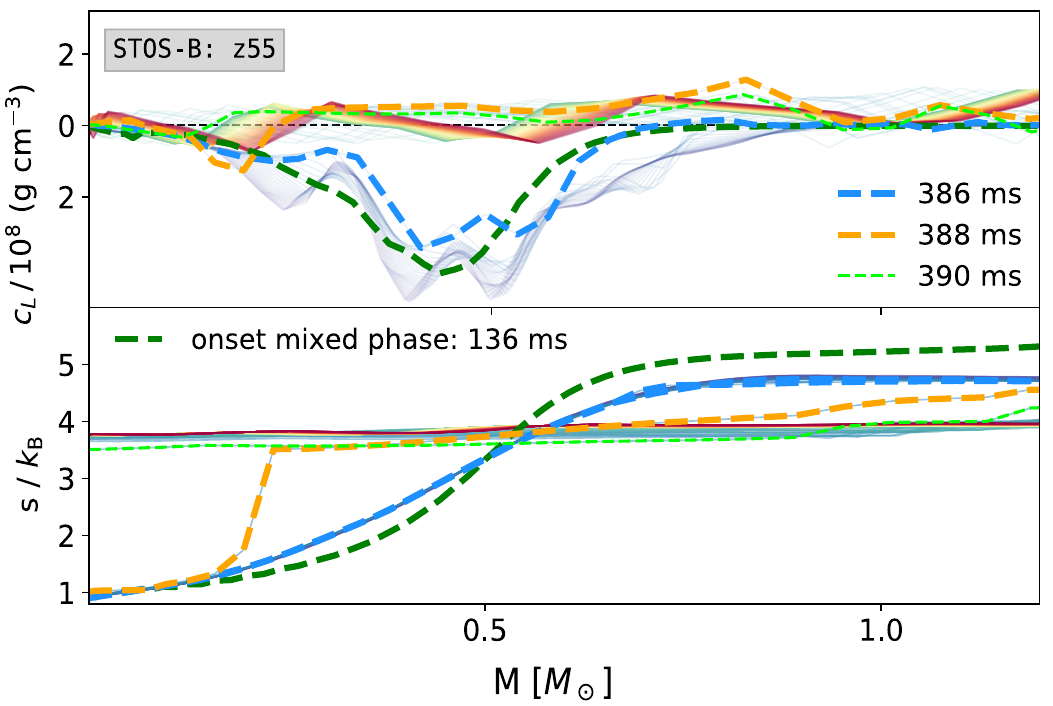}
	\caption{Relativistic Ledoux criterion
	(Equation~\ref{eq:ledoux_gr}, top) and specific entropy  (bottom) as function of enclosed mass around the time of the second collapse for STOS model \texttt{z55}.
	At $386\, \mathrm{ms}$ after the first bounce, the entropy profile still exhibits the typical post-bounce shape with a positive gradient out to $\mathord{\sim}0.7 \text{M}_\odot$ and a flat gradient in the proto-neutron star convection zone further outside. The Ledoux criterion shows a large negative value at the low-entropy core. Another $2\, \mathrm{ms}$ later, the
	Ledoux criterion has become positive in most
	of the formerly stable region inside $\mathord{\sim}0.7 \text{M}_\odot$, i.e., large parts of the PCS have entered the regime of inverted convection. The entropy gradient has been flattened considerably, but is still positive, and the innermost part of the core is not yet fully mixed. At $390\, \mathrm{ms}$, the innermost $\mathord{\sim}0.9 \text{M}_\odot$ of the PCS have formed on big convection zone with a flat entropy profile.\label{fig:inv_conv}}
\end{figure}

The increase (or even jump) in central entropy
during the mixed phase sets the STOS and CMF models apart from the DD2F\_SF models and may act against a precipitous collapse of the PCS core and a strong rebound.  This temperature increase in the STOS and CMF models during the mixed phase is a consequence of ``inverted convection''
inside the PCS during the phase transition, a phenomenon predicted by
\citet{Yudin2016-aw}.
As \citet{Yudin2016-aw} showed, the (non-relativistic) Ledoux criterion for convective instability
\begin{align}
\label{eq:ledoux}
    C_\mathrm{L}
    & =\dv{\rho}{r}
    -\frac{1}{c_\mathrm{s}^2}\dv{P}{r}
    >0,
    \intertext{can be expressed as}
C_\text{L} & = -\frac{c^2 T \rho}{P\Gamma c_V}
\left(\frac{\pd P}{\pd T}\right)_{\rho,Y_\mathrm{lep}}
\dv{s}{r}+
\left(\frac{\pd \rho}{\pd Y_\mathrm{lep}} \right)_{P\!,s}
\dv{Y_\text{lep}}{r} >0.
\end{align}
in terms of radial derivatives of entropy $s$ and lepton number $Y_\mathrm{lep}$ with the positive heat capacity $c_V= T (\pd S /\pd T)_\rho$. Since $(\pd P/\pd T)_{\rho,Y_\mathrm{lep}}$
is positive under normal conditions, a negative entropy gradient tends to be destabilising, but 
during a phase transition $(\pd P/\pd T)_{\rho,Y_\mathrm{lep}}$ can switch sign so that a positive entropy gradient (as usually encountered in the PCS core) acts as destabilising instead.
Qualitatively, the destabilisation during a phase transition can
also be understood from Equation~(\ref{eq:ledoux}).
A higher compressibility of mixed phase matter increases the compactness $\sim M/R$ of the PCS, leading to a steeper pressure gradient and an enlarged destabilising term $-c_\mathrm{s}^{-2} \pd P/ \pd r$. 
In the relativistic case \citep{Yudin2016-aw}, the situation
is qualitatively similar, but additional derivatives appear,
and the onset of inverted convection is not strictly limited
to $(\pd P/\pd T)_{\rho,Y_\mathrm{lep}}<0$. Furthermore, stabilisation or destabilisation by lepton number 
also needs to be taken into account so that
$(\pd P/\pd T)_{\rho,Y_\mathrm{lep}}<0$ is only an approximate condition for the onset of inverted convection, and the relativistic Ledoux criterion
\citep{thorne_1966} needs to be evaluated directly
\begin{equation}\label{eq:ledoux_gr}
    C_\text{L} = \dv{\rho(1+\epsilon/c^2)}{r} - \frac{1}{c_s^2}\dv{P}{r} > 0
\end{equation}
with the speed of sound $c_s^2 = \Gamma P / (\rho[1 + \epsilon/c^2 + P/(\rho c^2)])$.  
Figure~\ref{fig:inv_conv} shows $C_\mathrm{L}$
for model \texttt{z55} with the STOS EoS at selected
times around the time of the second collapse along with entropy profiles. As the PCS enters
the mixed phase, a wide unstable region appears
around $388\, \mathrm{ms}$ after the first bounce,
and a steep entropy gradient at the edge of the low-entropy core is largely eliminated. At
$388\, \mathrm{ms}$, a small positive entropy
gradient remains in the region where
Equation~(\ref{eq:ledoux_gr}) indicates instability,
hinting at the underlying phenomenon of inverted convection. Figure~\ref{fig:inv_conv} also shows that the mixing of the PCS core still takes a few
milliseconds, i.e., requires several dynamical timescales. 

\begin{figure}
	\centering
	\includegraphics[trim=10 5 10 14mm,
	width=\linewidth]{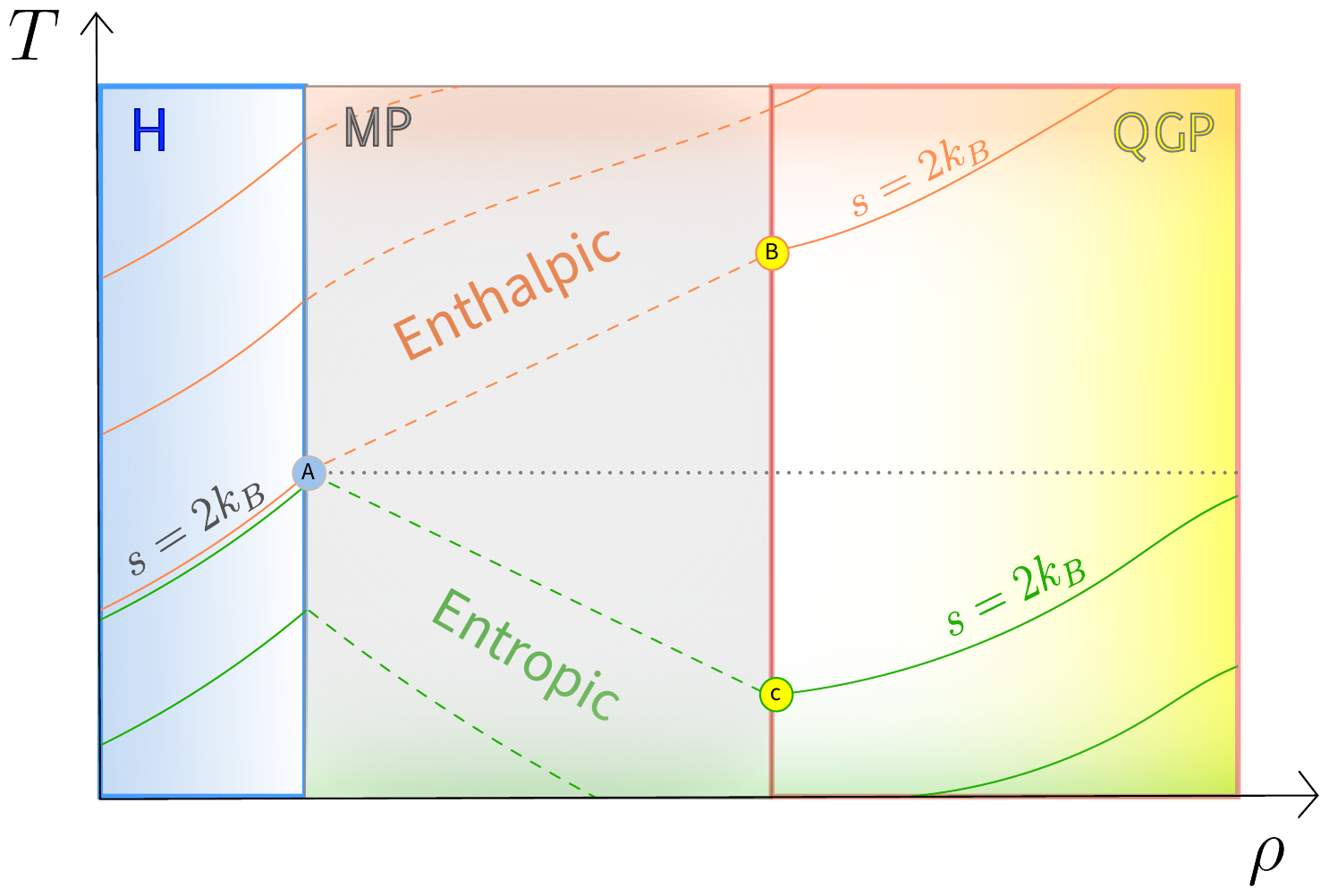}
	\caption{Schematic phase diagram with temperature as function of density for an enthalpic (orange) and entropic (green) EoS. All EoS in our setup feature entropic behaviour during during the mixed phase from hadrons to quarks. A mixed phase is entropic if the quark phase has higher entropy after isothermal compression (see grey dotted line). This implies a decrease in temperature for adiabatic compressions (a) $\rightarrow$ (c). For enthalpic PT the entropy decreases in the higher phase, which implies an increase in temperature for adiabtic compression (a) $\rightarrow$ (b).}
	\label{fig:illustration}
\end{figure}

There is in fact a deeper connection between
the properties of the phase transition and the
onset of inverted convection
\citep[see also][]{Yudin2016-aw,Hempel2017-ah}.
\citet{Iosilevskiy_undated-pk} classified first-order phase-transitions into two categories based on the sign of the change in enthalpy or second-order partial derivative of the Gibbs free energy in the mixed phase, i.e.,
\begin{align}
\Delta H < 0 &\rightarrow \left(\frac{\partial P}{\partial T}\right)_{\mathrlap{s}}\,>0 \,\,\,\text{(enthalpic)}\label{eq:enth} \\
\Delta H > 0 &\rightarrow \left(\frac{\partial P}{\partial T}\right)_{\mathrlap{s}}\,<0\,\,\,\text{(entropic)}\label{entr},
\end{align}
Here $\Delta H = T(S_\text{Quarks} - S_\text{Hadrons}$) is the enthalpy difference between the two phases before and after the phase transition. The underlying entropic nature of the hadron-quark phase transition thus can be understood as a consequence of higher entropy in the quark phase\footnote{The Clausius-Clapeyron relation~\citep{Landau1980-ea} is often referenced for the negative pressure gradient $\dv{P}{T}$ for Maxwellian phase transitions (local charge neutrality during the mixed phase) where pressure in the mixed phase is independent of density which is not the case for a Gibbs construction.}. When hadrons are deconfined into their constituents under adiabatic compression, the degrees of freedom become larger thus the kinetic energy per particle becomes smaller. The behaviour of entropic and enthalpic mixed phases is illustrated in Figure~\ref{fig:illustration}. 

Furthermore, the negative gradient of adiabats in the $P-T$-plane can be related to negativity of several other thermodynamic cross derivatives by simple Maxwell relations and particularly to the effect of ``thermal softening''~\citep[see, e.g.,][]{A_Steiner_M_Prakash_JM_Lattimer2000-qp,Iosilevskiy_undated-pk,Nakazato2010-yc,Hempel2013-su,Hempel2017-ah},
\begin{equation}
\left(\pdv{T}{\rho}\right)_{s} \leq 0
\Leftrightarrow  
\left(\pdv{P}{T}\right)_{\rho} \leq 0\label{eq:a}.
\end{equation}
The phase diagrams of all three EoS in Figures \ref{fig:dd2 phase}, \ref{fig:stos phase},
and \ref{fig:cmf phase} show a decrease in temperature during adiabatic compression in some parts of the mixed phase and crossover region, and
the effect is most pronounced for the DD2F\_SF EoS.
It is therefore somewhat puzzling that the DD2F\_SF  models do not show a similar amount of central heating by inverted convection as the STOS and CMF models. This unexpected finding can be resolved, however, by noting that the DD2F\_SF models still show some drift to higher core entropy (Figure~\ref{fig:dd2 phase}) during the phase transition, and that the impact of inverted convection also depends on the extent of unstable regions and the available time for mixing.

As discussed before, most of the STOS and CMF models do not evolve through the mixed phase rapidly, in contrast to the DD2F\_SF models.
This is because the strong entropic phase transition in DD2F\_SF is also linked to the \textit{thermal softening} (Equation~\ref{eq:a}) of matter, i.e., the decrease of
$\Gamma$, which is related to the behaviour of
$\pd P/ \pd T$ and $\pd P/\pd S$.
Entropic phase transitions, therefore, leads to a softer EoS region and is linked to the stability of stars~\citep{steiner_01}.
The different dynamics of the DD2F\_SF models as opposed to the STOS and CMF models could therefore be interpreted as follows: For a strong entropic phase transition that requires considerable latent heat to deconfine the quarks, significant softening leads to a rapid collapse on a dynamical timescale that leaves no time for significant mixing by convection and potentially results in a strong bounce. This hypothesis is substantiated by our results for high-compactness progenitors in the STOS setup (blue/green in Figure~\ref{fig:stos phase}) showing faster collapse times and lesser mixing in their cores (two of those models exhibiting a second core bounce and neutrino burst).  
On the other hand, for an entropic phase transition with small latent, the PCS can traverse the mixed phase more slowly, leaving a substantial fraction of its core concurrently in the unstable regime of the mixed phase for inverted convection to become effective, and mitigating the violence of the collapse and a (potential) rebound. We note in passing that the high specific entropies reached in the PCS due to inverted mixing are significantly larger than those reached in a recent study on simulating heavy ion collisions and binary neutron star mergers using the same underlying CMF EoS we used~\citep{most_2022}.

\subsection{Neutrino signals}
\label{sec:neutrino}
The neutrino signals of the exploding and non-exploding models show a variety of behaviours, including some features that have not yet been observed in models of PT-driven explosions.

\begin{figure*}
	\centering
	\includegraphics[trim=10mm 10mm 5mm 5mm, width=\linewidth]{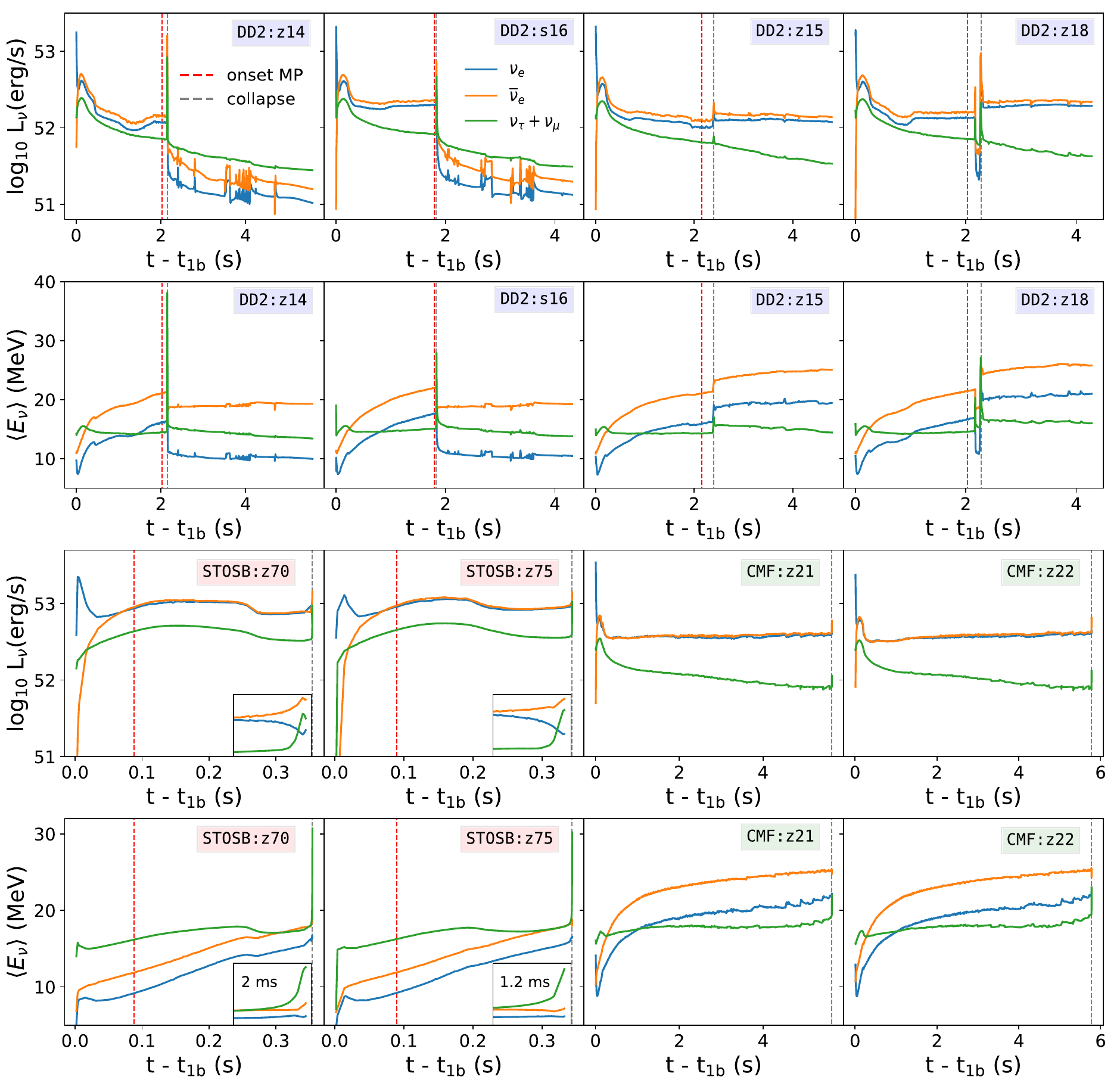}
	\caption{Neutrino luminosity and neutrino mean energy for $8$ selected progenitors as function of time after the first bounce. Vertical dashed lines show onset of mixed phase (dashed red) and collapse (dashed grey). The colour indicates the neutrino flavour. 
	The first and second row shows neutrino signals for the two exploding models \texttt{z14} and \texttt{s16} using DD2F\_SF (indicated by a blue box). \texttt{z15} and \texttt{z18} belong to the class of failed explosion where a second neurino burst is emitted without an explosion following. 
	Furthermore \texttt{z15} did not form a BH within the simulation time.
	The third row shows output for two progenitors in the STOS-B145 setup (indicated by a red box). 
	The two models \texttt{z70} and \texttt{z75} exhibit a PT-driven second neutrino burst before they collapse to a BH. The luminosities of these two models show a lower peak luminosity compared to the DD2F\_SF luminosities above.  The small inserts in the bottom right of the panels for models \texttt{z70} and \texttt{z75} magnify the last $2\,\mathrm{ms}$ and $1.2\,\mathrm{ms}$, respectively, before BH formation.  The $y$-axis in the inserts of Row 3 ranges from $52.5\,\mathrm{erg/s}$ to $53.2\,\mathrm{erg/s}$ (logarithmic scale); in Row 4, the $y$-axis of the insert ranges from $14\,\mathrm{MeV}$ to $31.9\,\mathrm{MeV}$ (linear scale).
	The fourth row shows the two progenitors \texttt{z21} and \texttt{z22} in the CMF setup (indicated by a green box). Both models collapse into BHs on a short timescale without experiencing a second core bounce or second neutrino burst. The rise in the neutrino signal is due to contraction of the neutrinosphere during collapse.\label{fig:lumi}}
\end{figure*}

We first consider the two exploding models \texttt{s16} and \texttt{z14} using the DD2F\_SF EoS.
Neutrino luminosities and mean energies for these two models are shown in Figure~\ref{fig:lumi}.
Immediately after the second collapse, the neutrino emission conforms to the behaviour known from previous works on PT-driven explosions.
 Once the shock wave reaches the close-by neutrino sphere, a second burst is released, which is dominated by electron antineutrinos \citep{Sagert2009-nd,Dasgupta2010-bz,Fischer2010-au,Fischer2012-pl}.
After the second neutrino burst, heavy flavour neutrinos exhibit higher luminosity than electron flavour neutrinos due to the quenching of accretion luminosity after successful shock revival. The lack of accretion luminosity is also responsible for the
drop in the mean energy of $\nu_\mathrm{e}$ and $\bar{\nu}_\mathrm{e}$ compared to the phase prior to the second collapse.

Later on, the luminosities $\nu_\mathrm{e}$ and $\bar{\nu}_\mathrm{e}$ show intermittent blips of enhanced neutrino emission. The temporal pattern of these blips appears chaotic, and they could be superficially dismissed as glitches in the neutrino transport solver. These are not unphysical, however, but related to the rather tepid nature of the explosions. Due to relatively slow 
expansion of the forward shock, the (reverse) termination shock of the neutrino-driven wind from the PCS forms at small radii (see, e.g., panel k in Figure~\ref{fig:radial profiles}). The reverse shock quickly starts to propagate backwards, and eventually reaches the PCS. The dense shell ahead of the reverse shock is accreted in the process, which gives rise to a transient enhancement of the emission of $\nu_\mathrm{e}$ and $\bar{\nu}_\mathrm{e}$ as accretion luminosity. The accretion is too weak to permanently stifle the outflow, and the neutrino-driven wind is reestablished, which again quenches the accretion luminosity. After a while, a reverse shock is formed again, and the same process repeats a few times at irregular intervals.
If observed, such an irregular enhancement of the electron flavour luminosity after a second burst would thus help to differentiate the energetics of PT-driven explosions.

For models that undergo a second bounce, but fail to explode, we find a similar variety of behaviours as \citet{Zha2021-xk}. Only a fraction of these also emit a second neutrino burst; in most cases the shock does not make it to the neutrinosphere.
For the DD2F\_SF EoS we find a second bounce for the majority of progenitors with the exception of  \texttt{z20}, \texttt{z21}, and \texttt{z22}. Out of all the DD2F\_SF models with a second bounce, however, only the models \texttt{z15}, \texttt{z18}, and \texttt{z19} emit a second neutrino burst.
In the STOS-setup only two progenitors with a high compactness parameter undergo a second core bounce. Both of these models emit a second neutrino burst dominated by $L_{\overline{\nu}_\mathrm{e}}$.
On the other hand, the CMF EoS does not lead to any 2nd core bounce and consequently no second neutrino burst either.  

Figure~\ref{fig:lumi} shows neutrino luminosities and mean energies for some of these non-exploding cases.
The failed explosion models \texttt{z15}, \texttt{z18}, and \texttt{z19} (DD2F\_SF EoS) 
are characterised by weaker second bursts than the exploding models \texttt{z14} and \texttt{s16}.
In \texttt{z15}, the  luminosities of
$\nu_\mathrm{e}$ and $\bar{\nu}_\mathrm{e}$ and the and mean energies
of all flavours settle at higher values than before the second collapse right after the burst due to the smaller radii and higher temperatures of the neutrinospheres. The failed explosion model \texttt{z18} shows a precipitous drop of electron flavour luminosities and mean energies right after the burst due to the transient quenching of accretion and then another burst before the electron flavour luminosities and all the mean energies settle at higher values than before the second collapse. This behaviour was already seen by \citet{Zha2021-xk} in their simulations. The second peak is stronger with $L_{\overline{\nu}_\mathrm{e}}\approx 5\times 10^{52} \, \mathrm{erg}\, \mathrm{s}^{-1}$ compared to $L_{\overline{\nu}_\mathrm{e}}\approx 1\times 10^{53}\, \mathrm{erg}\, \mathrm{s}^{-1}$.

The two STOS models with a second bounce (\texttt{z70} and \texttt{z75}) both emit a second neutrino burst right before black hole formation. The second neutrino burst is dominated by $\overline{\nu}_\mathrm{e}$, followed by $\nu_{\tau,\mu}$. It is not immediately obvious from Figure~\ref{fig:lumi} whether the rather small peak in $L_{\nu_\mathrm{e}}$ and $L_{\overline{\nu}_\mathrm{e}}$ is indeed from the shock breakout or from increasing temperature of the collapsing neutrinosphere during the collapse to black hole. In the inset at the bottom right for both progenitors we therefore show the last $1.2 \mathrm{ms}$ (\texttt{z70}) and $2 \mathrm{ms}$ (\texttt{z70}). We find an increase (decrease) of the $\overline{\nu}_\mathrm{e}$ ($\nu_\mathrm{e}$) luminosity before collapse, i.e., there is indeed a second burst from shock breakout that is not connected to black hole formation per se. The slightly delayed, more sudden increase in $\nu_{\tau/\mu}$ results from the contraction of the neutrinosphere in the course of black hole formation. This effect also explain why (different
from the DD2F\_SF models) the heavy flavour neutrinos reach higher mean energies
than the electron flavour neutrinos during
the second burst.

The four bottom right plots show the progenitors \texttt{z21} and \texttt{z22} in the CMF setup. Those progenitors do not undergo a second core bounce and immediately form BHs once they enter the crossover region. The slight increase in the neutrino luminosity (and mean neutrino energy) is due to the contraction of the neutrino-sphere which heats the material and increases neutrino emission rates. The hierarchy is similar to DD2. 

\subsection{Nucleosynthesis in the Ejecta}
CCSNe are one of the main production sites for heavy elements in the universe. There are, however, still many open questions about the contribution of CCSNe to heavy elements beyond the iron group. The long-standing notion that CCSNe are the dominant source of rapid neutron-capture process (r-process) elements has been revised in recent years both because modern simulations neither showed the requisite high entropy and low electron fraction in the neutrino-driven wind \citep{huedepohl_10,fischer_10}, and because compact binary mergers were identified as a robust site for the r-process \citep[e.g.,][]{freiburghaus_99,goriely_11}. It remains conceivable though, that CCSNe contribute to some r-process production, especially in low-metallicity environments \citep{spite_1978,cowan_1995,hansen_2014}. Jet-driven explosions have been considered
extensively as an r-process site \citep{nishimura_06,winteler_12,moesta_18,grimmett_21}. PT-driven explosions are also a possible candidate because the rapid ejection of material from the PCS surface may provide the requisite neutron-rich conditions for an r-process. Based on 1D explosion models, \citet{Fischer2020-rg} found considerable amounts of r-process material, reaching up to and beyond the third peak. 
The yields from PT-driven explosions are also of interest
because they may potentially provide constraints on the
rates or existence of these events.

\begin{figure}
  \centering
  \includegraphics[width=0.9\linewidth]{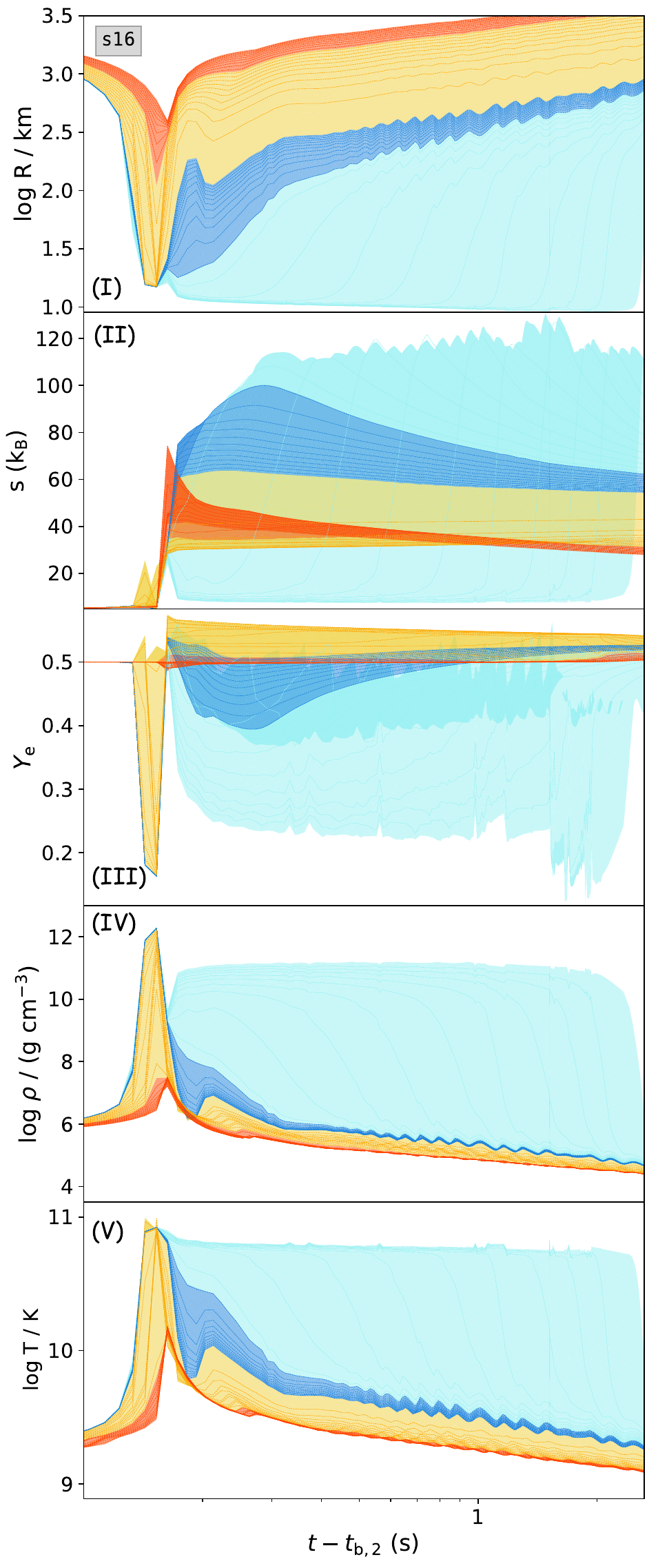}
  \caption{Radial trajectories $R(t)$, entropy $s$, electron fraction $Y_\mathrm{e}$, density $\rho$, and temperature $T$ of selected ejected mass shells as function of logarithmic time (s) after the second bounce for the exploding model \texttt{s16}. Matter can be distinguished into (I) prompt ejecta (orange), (II) early neutrino-processed ejecta (yellow),  (III) intermediate ejecta (blue), and (IV) neutrino-driven wind material (turquoise).}
  \label{fig:mass trajectories s16}
\end{figure}

We identify ejecta based on the same criterion as for the calculation of the diagnostic explosion energy. A mass shell is \emph{ejected} if the local specific binding energy and radial velocity are both positive. 
Figure~\ref{fig:mass trajectories s16} shows
the (I) radius $R$, (II) entropy, (III) electron fraction $Y_\mathrm{e}$, (IV) density $\rho$, and (V) temperature for selected ejecta trajectories as functions of time after the second bounce for the exploding model \texttt{s16} using the DD2F\_SF EoS. The ejecta depicted in Figure~\ref{fig:mass trajectories s16} cover material in the neutrino-driven wind from
mass coordinate $1.7585\,\mathrm{M}_\odot$ out to shells 
starting from a few thousand kilometres that were shocked early after the second bounce. The Figure does not include all the
ejecta that are shocked later after the second bounce.
Following \citet{Nishimura2011-yo}, we classify ejected matter into:
\begin{enumerate} 
\item Prompt ejecta at larger distance $R \gtrsim 100\,\mathrm{km}$ from the PCS (orange),
\item Ejecta exposed to $\nu$-heating with $Y_\mathrm{e}\geq 0.5$  which initially stall after the second core bounce and later get ejeteced due to neutrino heating (``Early neutrino-processed ejecta'', yellow),
\item Intermediate ejecta exposed to $\nu$-heating with $Y_\mathrm{e}\leq 0.5$  (blue),
\item Wind ejecta (turquoise) (turquoise).
\end{enumerate}
Prompt ejecta (orange) are expelled by the shock wave with no contact to the neutrino-driven wind. At collapse time, they fall in to radii $R\approx 100\, \mathrm{km}$~(see I) before they get expelled and keep their initial $Y_\mathrm{e}\approx 0.5$ from the infall phase. 
In some of these trajectories, $Y_\mathrm{e}$ decreases slightly below $0.5$ due to antineutrino captures during the second burst.
The early neutrino-processed ejecta (yellow band) are initially propagating outwards after the 2nd collapse, followed by a transient phase of negative velocity before they get re-accelerated by neutrino energy disposition and mechanical work by the wind material inside. 
Before these ejecta are hit by the second shock,
they come close enough to the PCS to reach a relatively low electron fraction $0.50\gtrsim Y_\mathrm{e}\gtrsim 0.18$ as a consequence of electron captures.
When hit by the second shock, ejection is not fast enough to conserve the low $Y_\mathrm{e}$, and the electron antineutrino burst has no substantial influence on the final $Y_\mathrm{e}$ because it occurs when the early neutrino-processed ejecta are still at high densities and charged-current processes have not frozen out yet. Instead, the freeze-out value of $Y_\mathrm{e}$ is determined by the neutrino emission during the rapid decay phase of the neutrino luminosities after the second burst. At this early time, the relative difference
between electron neutrino and antineutrino luminosities and mean energies is still small enough to make the ejecta proton-rich, analogous to the situation in neutrino-driven explosions \citep{pruet_05,froehlich_06}.
The intermediate ejecta (blue), like the early neutrino-processed ejecta, are driven outwards initially by the second core bounce before they briefly fall in once more to about $20\, \mathrm{km} \gtrsim R \gtrsim 100\, \mathrm{km}$ (see I) and deleptonise to moderately low $Y_\mathrm{e}$ before they are ejected. 
As a consequence of a slightly increasing difference between
electron neutrino and antineutrino mean energies, neutrino processing slowly raises $Y_\mathrm{e}$ during ejection. On the other hand, ejection is rather slow and neutrino-processing brings this ejecta component
to $Y_\mathrm{e}\approx 0.5$.
The $Y_\mathrm{e}$ and entropy in these ejecta are, due to their small amount of mass, subject to a small drift later on because of numerical diffusion and trajectory integration inaccuracies.
The mass shells in the neutrino-driven wind (turquoise) are exclusively unbound by neutrino heating. The wind is characterized by moderately high entropies of up to
$120 \, k_\mathrm{B}/\mathrm{baryon}$. 
The excess of electron antineutrino emission over electron neutrino emission for an extended period sets favorable conditions for low $Y_\mathrm{e}$ in the first few seconds after the second collapse. These conditions reverse again later
when the relative difference between the electron antineutrino and neutrino luminosity decreases again  (see graph for model \texttt{s16} in Figure~\ref{fig:lumi}
at $\mathord{\sim}3.5\,\mathrm{s}$). Hence neutrino heating turns the neutrino-driven wind slightly proton-rich with $Y_\mathrm{e} \approx 0.52$ at later times $t_\mathrm{2,c}\gtrsim 2\,\mathrm{s}$.

Our findings of early neutron-rich neutrino-driven ejecta for model \texttt{s16} which, at later times, turns slightly proton-rich were also found (in the context of hybrid EoS) in~\citet{Nishimura2011-yo}.

\begin{figure}
  \includegraphics[trim=10mm 10mm 5mm 5mm, width=\linewidth]{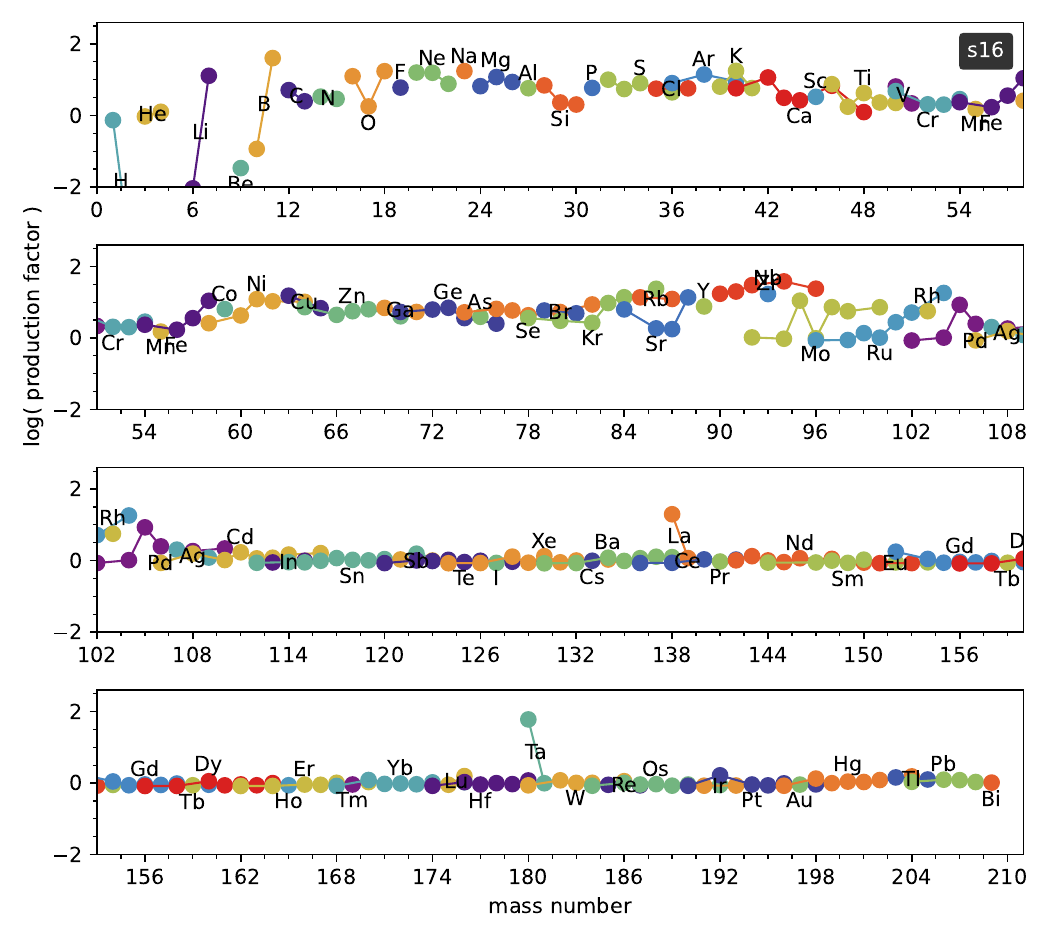}
   \caption{Logarithmic isotopic production factors $\log \left(X_i / X_{i,\odot}\right)$ for model \texttt{s16}. Edges connect isotopes which belong to the same element. The two isotopes $^{94}$Zr (production factor 38) and $^{11}$B (production factor 40) show the highest overproduction factor. $^{139}$La and $^{180}$Ta are produced after the first core bounce and before the second core bounce by neutrino process in the outer shells.}
   \label{fig:s16 prodfac}
\end{figure}%
\begin{figure}
  \includegraphics[trim=10mm 10mm 5mm 5mm, width=\linewidth]{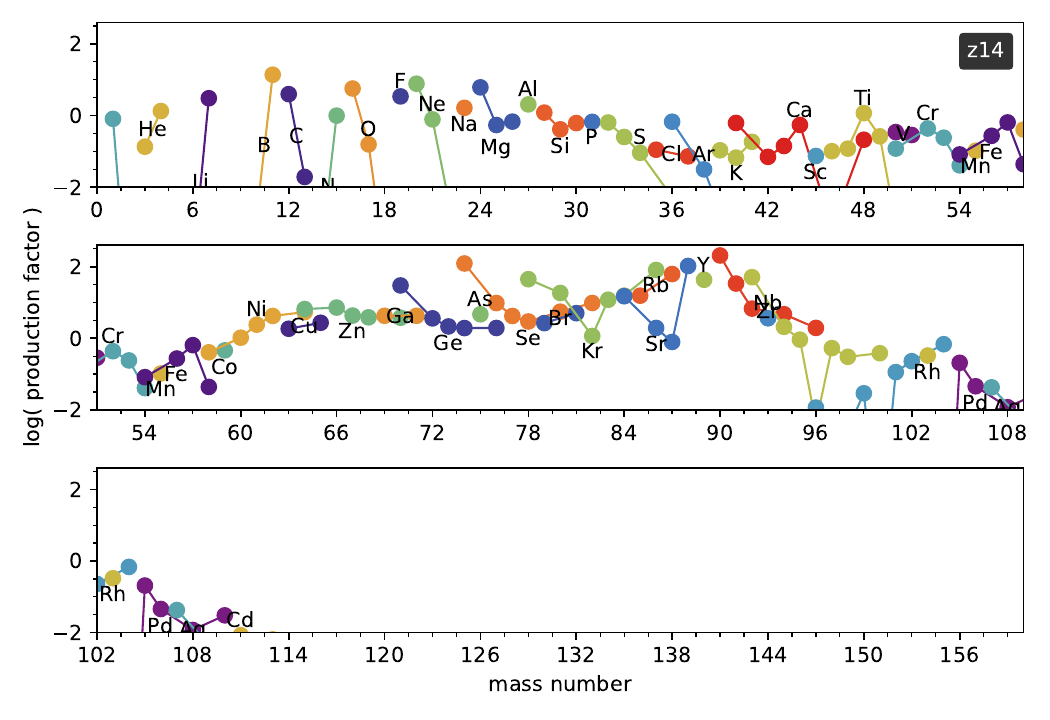}
  \caption{Logarithmic isotopic production factors $\log \left(X_i / X_{i,\odot}\right)$ for model \texttt{z14}. Edges connect isotopes which belong to the same element. The two isotopes  $^{90}$Zr (production factor 200) and $^{11}$B (production factor 13) have the highest overproduction factors.}
  \label{fig:z14 prodfac}
\end{figure}        
\noindent

In Figure~\ref{fig:s16 prodfac} and~\ref{fig:z14 prodfac} we plot isotopic production factors for models \texttt{s16} and \texttt{z14}, respectively. To better understand the nucleosynthesis processes that shape the yield pattern, we also consider spatially-resolved information. 
Figure~\ref{fig:plot_map_z14} shows the abundances of decayed nuclei as function of logarithmic mass coordinate (counting from the mass cut) for \texttt{s16} (upper plot) and \texttt{z14} (lower plot). These clearly reflect the different nucleosynthesis regimes that shape the salient features of the overall yields.
\subsubsection{Iron Group}
\begin{figure}
\centering
\begin{minipage}{1.0\linewidth}
\centering 
\includegraphics[trim=10mm 10mm 5mm 5mm, width=1.\linewidth]{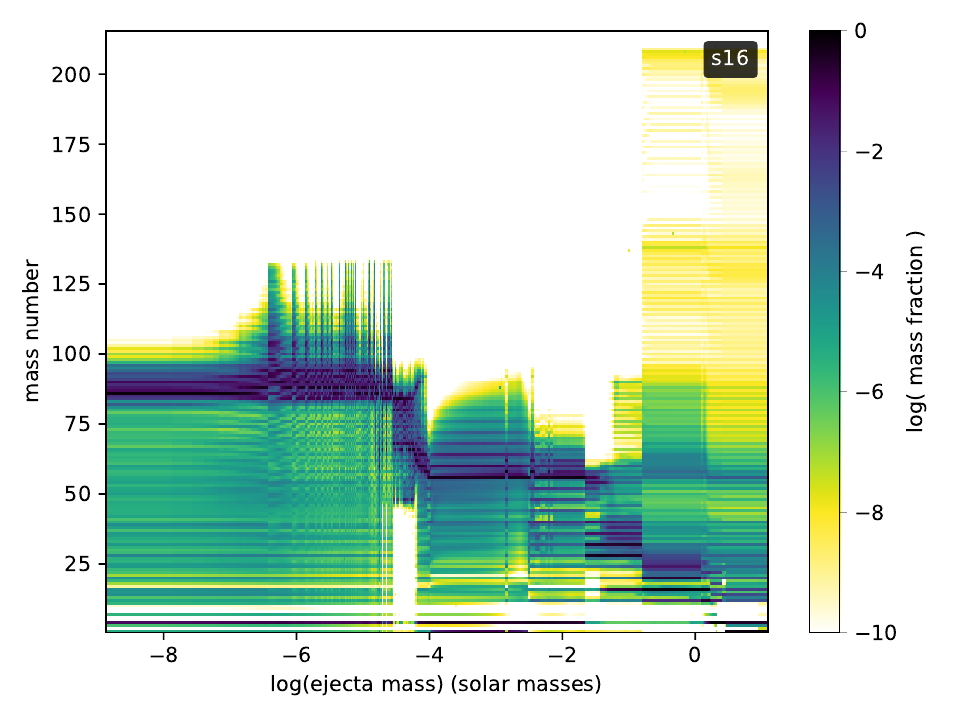}
\includegraphics[trim=10mm 10mm 5mm 5mm, width=1.\linewidth]{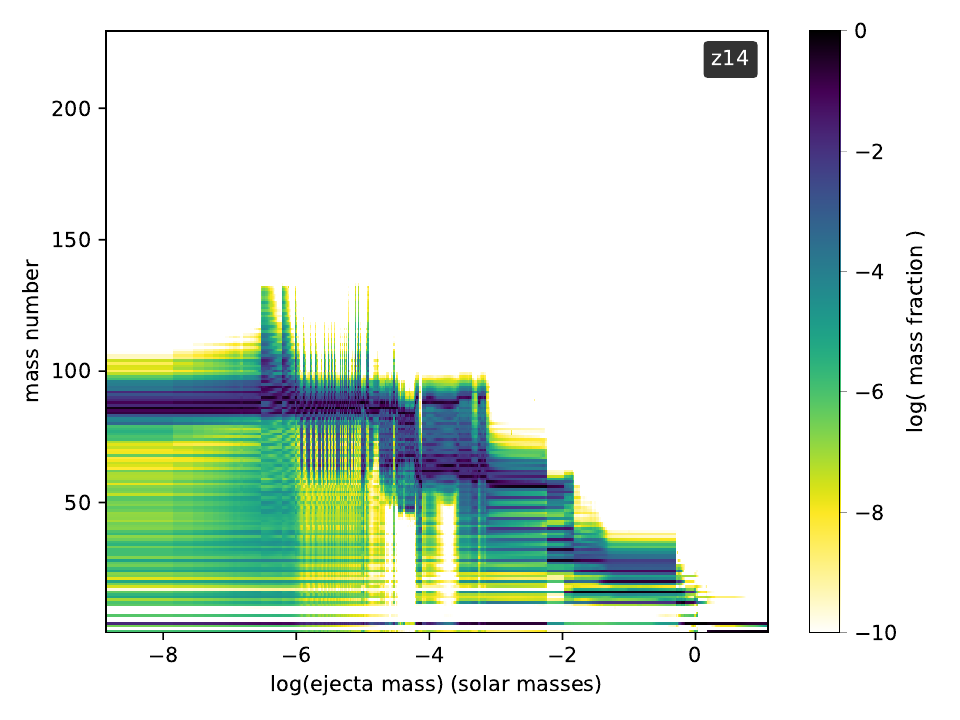}
  \caption{Mass numbers $A$ of ejected material for the exploding progenitors \texttt{s16} (upper panel) and \texttt{z14} (lower panel) as function of log ejected mass, colour coded by the abundance.\label{fig:plot_map_z14}}
\end{minipage}
\end{figure}

The iron group elements around $A\sim 56$ are produced in regions with $Y_\mathrm{e}\approx 0.5$ at moderate entropy $s\sim 40 k_\mathrm{B}/\mathrm{nuclon}$  within the prompt ejecta (orange band in Figure~\ref{fig:mass trajectories s16}). 
They originate from mass coordinates $10^{-4}\,\text{M}_\odot$ to
about $10^{-2}\,\text{M}_\odot$ outside the mass cut, seen as dark blue bands for both progenitors in Figure~\ref{fig:plot_map_z14}.
Due to the thermodynamic conditions in this region, the
yields are determined by normal or $\alpha$-rich freeze-out from NSE. In fact, two different regimes can clearly
be distinguished in 
the upper plot of Figure~\ref{fig:plot_map_z14} for model \texttt{s16}, with the inner region showing stripe-like patterns of relatively high mass fractions above $A=56$ due to $\alpha$-captures.
In the case of model  \texttt{z14} (Figure~\ref{fig:plot_map_z14}), conditions around the mass coordinate $10^{-4} \text{M}_\odot$ also permit iron-group nucleosynthesis, but light-particle capture
reactions play a greater role and lead to a more significant production of trans-iron elements as we shall discuss below.
We tabulate individual decayed stable iron group nuclides in Table~\ref{table:lepp table}. 
The total ejected mass of those iron group nuclides with $45\leq\text{Z}\leq 56$ is $3.623\times 10^{-2} \text{M}_\odot$ for
model \texttt{s16}  and  $4.960 \times 10^{-3} \text{M}_\odot$ for \texttt{z14}. 

\subsubsection{Light Element Primary Process}
The process leading to elements beyond iron with $A < 130$ is usually referred to as Light Element Primary Process (LEPP; \citealt{Travaglio_2004,2011ApJ...731....5A}). 
LEPP isotopes in the upper plot of Figure~\ref{fig:plot_map_z14} are produced in neutron-rich intermediate ejecta (blue band in Figure~\ref{fig:mass trajectories s16}).  The two LEPP-process zirconium isotopes
$^{90}$Zr (model~\texttt{z14}) and $^{94}$Zr (model~\texttt{s16}) show significant overproduction factors. Their absolute yields are tabulated along other LEPP process isotopes in Table~\ref{table:lepp table}. 
\begin{table}
\input{tables/lepp.tex}
\caption{Left: Absolute masses (solar mass) for (stable decayed) LEPP material for the exploding models \texttt{z14} (second column) and \texttt{s16} (third column). Right: Absolute masses (solar mass) for (stable decayed) iron group nuclides. }
\label{table:lepp table}
\end{table}

\subsubsection{Neutrino process in the envelope:  $^{11}$B, $^{138}$La, and $^{180}$Ta}
Model \texttt{s16} exhibits significant overproduction of
 the heavy odd–odd nuclides $^{138}$La and $^{180}$Ta which are among the rarest
solar system species~\citep{Martinez-Pinedo:2017ksl}. 
We find overproduction factors of $\sim 60$ for $^{180}$Ta and $\sim 20$ for $^{138}$La for the exploding model \texttt{s16} whose production is however unrelated to the PT-driven explosion mechanism.

These nuclei belong to the neutron-deficient class of nuclei, called  $p$-nuclei~\citep{1995A&A...298..517R}. One likely origin for
these nuclei is the neutrino ($\nu$)-process \citep{Woosley:1989bd, 2005PhLB..606..258H}, which entails charged-current or neutral-current spallation processes induced by supernova neutrinos. This process also operates in model \texttt{s16}
and can be identified as the source of $^{138}$La and $^{180}$Ta since both isotopes are  produced after the neutrino burst from the first core bounce\footnote{The
$p$-process (also called $\gamma$-process) can be excluded since no shock has passed through the material at this point~\citep{Kusakabe_2010}.}.
$^{138}$La and  $^{180}$Ta are mainly produced by the charged reactions $^{138}\text{Ba}(\nu_{\mathrm{e}},e^-)\text{La}$ and $^{138}\text{Hf}(\nu_{\mathrm{e}},e^-)^{180}\text{Ta}$~\citep{Woosley:1989bd,2005PhLB..606..258H}. The first neutrino burst, dominated by $\nu_{\mathrm{e}}$, provides favorable conditions to these charged current reactions. 

We do not see $^{138}$La and $^{138}$Ta production for model~\texttt{z14} since the parent weak s-process isotopes $^{138}$Ba and $^{138}$Hf are not present in the zero metallicity model \texttt{z14}. 

It should be noted, however, that the $^{180}$Ta yield for \texttt{s16} is possibly lower than what our network predicts. The network calculates the sum of isomeric and ground states which have both very distinct half-lives (the lifetime of the  ground state is much shorter\footnote{An isomeric state is an exited meta-stable ($\tau \leq 1\,$ns than the ground state) nuclear state~\citep{2021ApJS..252....2M}.}).  This leads to an overestimation of $^{180}$Ta by a factor $0.3-0.5$~\citep{2002ApJ...576..323R}. Since $^{180}$Ta and $^{138}$La are both produced after the first neutrino burst and before the phase transition induced explosion, the EoS would have little effect on their production.  Additional and more detailed discussion about the interesting isotopes $^{138}$La and $^{180}$Ta can be found here~\citep{ARNOULD20031,PhysRevLett.98.082501,Austin:2011gw,lahkar,2018ApJ...865..143S}.
$^{11}$B is another rare $\nu$-process isotope that shows large production factors for both progenitors in Figure~\ref{fig:s16 prodfac} and~\ref{fig:z14 prodfac}. Contrary to the heavier $p$-nuclei isotopes $^{138}$La and $^{180}$Ta it is mainly produced via the neutral current $^{12}\text{C}(\nu,\nu' p)^{11}\text{B}$ on the rather abundant $^{12}$C~\citep{2014arXiv1412.8425L}. For both progenitor models it is produced after the second core bounce and subsequent neutrino burst. 
\subsubsection{Rate constraints}
The overproduction factors can in principle be used to constrain the fraction of PT-driven explosions (of a similar type as \texttt{s16} and \texttt{z14}) among all CCSNe by assuming that the most overproduced  nucleus solely originates from this particular type of explosion \citep[e.g.,][]{Wanajo:2009igs}. Based on Figure~\ref{fig:s16 prodfac} and \ref{fig:z14 prodfac}, the nuclides with the largest production factor are  $^{94}$Zr for \texttt{s16} and $^{90}$Zr for \texttt{z14} .  We assume that there is a negligible contribution to galactic $^{16}$O production for each of these two representative PTD-CCSN and negligible contribution from non-PT driven CCSN to the production of the two zirconium isotopes $^{90}$Zr and $^{94}$Zr.  An approximate upper rate constraint on the fraction $f$~\citep{Wanajo:2009igs,Wanajo:2010ig,Wanajo:2017cyq} of PTD-CCSNe is then given by:
\begin{equation}
    \frac{f_\mathrm{s16/z14}}{1-f_\mathrm{s16/z14}} = \frac{X_\odot(^{A}\text{Zr})/X_\odot(^{16}\text{O})}{M(^{A}\text{Zr})/M_\text{CCSN}(^{16}\text{O})}\,\,
\end{equation}
where $f_\mathrm{s16/z14}$ stands for the ratio of the respective events to all CCSN-events. $A$ denotes the mass number of the Zr isotope. The nuclear reaction network yields an absolute mass for both isotopes of $M_\text{z14}(^{90}\text{Zr})=3.856\times 10^{-4}$ M$_\odot$ and $M_\text{s16}(^{94}\text{Zr})=2.637\times 10^{-6}$~M$_\odot$. $M_\text{CCSN}(^{16}\text{O})=1.5$ $\text{M}_\odot$ is the average mass production of $^{16}$O by massive CCSN with masses ~$M\in [13-40]\text{M}_\odot$ averaged over the stellar initial mass function~\citep{2006NuPhA.777..424N}. The other parameters are $X_\odot(^{16}\text{O})=7.377\times 10^{-3}$, $X_\odot(^{90}\text{Zr})=1.404\times 10^{-8}$ and $X_\odot(^{94}\text{Zr})=4.955\times 10^{-9}$ and taken from \citet{2021SSRv..217...44L}. 
The fraction of these kinds of explosions under the above assumptions is only constrained to be $f_\text{s16} \le 0.520$ for \texttt{s16} and to $f_\text{z14} \le 0.007$.
Thus, because of the large variations in yields between these two PT-driven explosions, no meaningful constraint can be placed based on the frequency or existence of PT-driven explosion yet.
Nucleosynthesis from individual PT-driven explosion models is clearly not sufficiently robust for this purpose. Integrating PT-driven explosions into the bigger picture of chemogalactic evolution would clearly require fine representative sampling of this hypothetical explosion channel by many models, which is not realistic at this stage given the current uncertainties on the high-density equation of state.

\section{Summary and Outlook}
\label{section four}
In this paper, we studied the effects of the QCD phase transition in CCSNe in order to better gauge the robustness and sensitivities of the proposed phase-transition driven explosion mechanism. We performed spherically symmetric CCSN simulations for up to 40 progenitors in the mass range $14\texttt{-}100\,\text{M}_\odot$ with three different hadron-quark EoS using the
general relativistic \textsc{CoCoNuT-FMT} code with an effective 1D mixing-length treatment for convection. We considered two hybrid EoS with a first-order phase transition to quark matter: DD2F\_SF contains stiffening repulsive quark interactions in the quark phase \citep{Fischer2017lag,Bastian2021-ga}; the STOS-B145 is matched to the Bag model in the quark phase \citep{Sagert2010-yw,Sagert2009-nd}.  The third EoS is a chiral mean-field model with a smooth crossover to quarks (CMF, \citealp{Motornenko:2019arp}).  

While phase-transition driven explosions have been proposed as a scenario for hyperenergetic supernovae from massive progenitors \citep{Fischer2017lag}, our results present a more nuanced view on the robustness and potential signatures of such explosions. We find
only two explosions among our large set of models,
namely for a $14\,\text{M}_\odot$ zero-metallicity
progenitor and a $16\,\text{M}_\odot$ solar-metallicity progenitors, both using the DD2F\_SF EoS. In another three  progenitors models (\texttt{z15}, \texttt{z18}, \texttt{z19}), the shock launched by the rebound after the phase-transition
induced collapse propagates out for several $100\, \mathrm{km}$, but ultimately fails to explode the star. 
There are different parameterizations available for the DD2F\_SF-EoS, which differ in various parameters for the treatment of hadronic and quark matter, such as the onset density of the mixed phase and the latent heat; these in turn influence the maximum cold neutron star mass and the mass-radius relation (see \citealp{Kaltenborn2017-ch,Bastian2021-ga} for a detailed discussion). The parameterization we chose has a higher onset density and slightly softer hadronic EoS compared to the one utilized in~\citet{Fischer2017lag,Fischer2020-rg,Fischer2021-zh}. For direct comparison, we simulated a small set of three progenitors using the DD2F\_SF parameterizations in~\cite{Fischer2017lag}, i.e., DD2F\_SF RDF~1.2, (which uses an excluded volume version of DD2F in the hadronic phase). We found very similar outcomes to DD2F\_SF RDF~1.4, with the explosion of \texttt{z14} having a low explosion energy $E_\mathrm{exp} \sim 10^{50}\,\mathrm{erg}$, BH formation for \texttt{z85}, and a failed explosion for \texttt{z15}. We conclude that the general trend we see towards low explosion energies (for exclusively low compactness models) is robust within small parameter variations in the DD2\_SF-EoS set for simulations with our \textsc{CoCoNuT-FMT} code. We leave a more systemic evaluation for the entire DD2F\_SF parameter-set open for future studies.
We find no successful explosions for the STOS-145 EoS, although two zero-metallicity progenitors of $70\,\text{M}_\odot$ and $75 \text{M}_\odot$ undergo a rebound after the phase-transition collapse and end up as ``failed explosions''. None of the models using the CMF EoS explode. Some low-compactness progenitors do not reach the phase transition at all, whereas models with moderate or high compactness quietly cross the phase transition without a second bounce and eventually collapse to black holes.

The higher likelihood of phase-transition driven second core bounces for low compactness for the DD2F\_SF EoS is
consistent with trends identified by \citet{Zha2021-xk}, (who applied the STOS-B145 EoS). The phase transition is reached significantly below the maximum stable TOV mass after the phase transition for the lower specific core entropy of the PCS in low-compactness models and a direct collapse to a black hole can be avoided. 
For a second bounce to occur, the maximum mass of the twin star branch in the mass-radius relation has to be larger than the entropy-dependent maximum mass in the large-radius branch \citep{Zha2021-xk}.

The collapse and rebound must also be sufficiently violent in order for an explosion to occur. In the lowest compactness parameter models, the central density of the PCS only increases modestly before a rebound occurs, and the resulting shock wave is too weak to trigger an explosion. This leaves only a small window for successful explosions at low, but not excessively low compactness.

Inspecting the phase diagrams of the different
EoS suggests that the structure of the phase transition for the DD2F\_SF EoS, with a strong softening (as reflected by the drop in the adiabatic index $\Gamma$) during the mixed phase and stiffening in the quark phase due to repulsive vector repulsion, provides more
favorable conditions for an explosion.  By contrast, the STOS EoS shows gradual softening towards higher densities, and the situation for the CMF EoS is similar for the relevant PCS core entropies. In this context it is important to note that equations of state that include any significant repulsive interaction, or stiffening, in the quark phase usually violate constraints from lattice-QCD simulations \citep{Steinheimer_2014} as well as the expected Stefan-Boltzmann limit at high densities.

A further dynamical difference between the DD2F\_SF models and the STOS and CMF models lies in the occurrence of ``inverted convection'' \citep{Yudin2016-aw}. During the phase transition, the positive entropy gradient in the PCS core becomes convectively unstable  due to the anomalous behaviour
of the thermodynamic derivative $(\pd P/\pd T)_\rho$, which can be connected to the entropic nature of the phase transition. As a result, the core of PCS is thoroughly mixed, and the centre of the PCS is heated significantly, raising the central entropy to about $4k_\mathrm{B}/\mathrm{baryon}$. The development of inverted convection could have a further damping effect on the collapse and rebound (where applicable) in the STOS and CMF models. For the DD2F\_SF EoS, conditions for inverted convection are met in principle, but due to the stronger softening in the mixed phase, the collapse and rebound occur too rapidly to allow significant mixing by convection in the PCS core.

Our neutrino signals largely show similar characteristic signatures of a first-order phase transition as in previous studies \citep{Sagert2009-nd,Zha2021-xk}. For successful or ``failed explosions'' (with transient outward propagation of the shock), there is a strong antineutrino burst \citep{Sagert2009-nd,Dasgupta2010-bz,Fischer2010-au,Fischer2012-pl}. Similar to \citet{Zha2021-xk}, some failed explosions are characterised by a short phase of strongly reduced neutrino emission followed by another burst when the shock stalls again and accretion onto the PCS recommences. For our two exploding models, we find a new neutrino signature, namely a series of mini-bursts during the explosion phase, which are the result of occasional fallback from the weak wind outflow from the PCS.

The dynamics of the two exploding models is markedly different from earlier studies \citep{Fischer2020-rg} that proposed PT-driven supernovae as a scenario for hyperenergetic explosions of massive progenitors. The two successful explosion models \texttt{s16} and \texttt{z14} only reach energies of
$\mathord{\sim}1.25\times 10^{50}\,\mathrm{erg}$ and $3.64\times10^{49}\,\mathrm{erg}$, respectively. The nucleosynthesis is also less spectacular than in previous studies that found significant r-process production \citep{Nishimura2011-yo,Fischer2020-rg}.
The models show appreciable overproduction of
some LEPP nuclei, in particular for model \texttt{z14}. The largest production factors
are found for $^{90}\mathrm{Zr}$ for model \texttt{z14} and for $^{94}\mathrm{Zr}$ for model \texttt{s16}. However, the differences between the two models indicate that PT-driven explosions would not be a very robust site for LEPP nucleosynthesis.
For this reason, chemogalactic evolution cannot meaningfully constrain the rate of putative PT-driven explosions.

On a more fundamental level, our results cast doubts on the viability of PT-driven explosions. Since conditions for a PT-driven explosion appear most favorable in low-compactness progenitors as for our DD2F\_SF models, it is likely that a neutrino-driven explosion would develop before the PCS enters the phase transition.
However, the progenitor dependence seen for the DD2F\_SF EoS may not be generic as indicated by the STOS models, where a second bounce does occur (albeit without triggering in explosion) for two high-compactness progenitors that likely would not explode by the neutrino-driven mechanism or at best experience an ``aborted explosion'' with massive fallback and black hole formation.

A more serious problem lies in the thermodynamic properties of the EoS and the nuclear physics assumptions that are required for successful explosions in our study. The explosions of the two DD2F\_SF models hinge on the combination of significant softening during the mixed phase followed by abrupt and pronounced stiffening in quark phase. This peculiar behaviour of the DD2F\_SF EoS rests on
the problematic assumption of vector repulsion in the quark phase. Although repulsive quark vector interactions have been used in the literature and fulfil the two solar mass constraint \citep{Kaltenborn2017-ch,Bastian2021-ga}, QCD lattice calculations disfavour repulsion among quarks in the other regimes of QCD at high temperatures and vanishing baryon density \citep{Steinheimer_2011, Steinheimer_2014}. 

These problems do not yet spell the end for a dynamical role of the hadron-quark phase transition in CCSNe, however. Further work is still necessary to resolve the tensions between different results on the progenitor dependence of PT-driven explosions. But even if the phase transition does not provide a pathway to a distinct explosion mechanism in its own right, its effects on CCSN dynamics and observables still deserve more nuanced attention. In particular, the phenomenon of inverted convection needs to be investigated in multi-dimensional simulations. First two-dimensional simulations of PT-induced collapse with a hybrid STOS-bag EoS (with a bag constant of $165\, \mathrm{MeV}$) were already presented by \citet{Zha2020-rv}, who found a prominent high-frequency gravitational wave signal during the phase around the second bounce due to convective activity deep inside the PCS. While \citet{Zha2020-rv} associated the new gravitational wave signature with the second collapse and bounce in the wake of a first-order phase transition, we speculate that such bursts could be more generic signatures of the hadron-quark phase transition as long as they provide conditions for inverted convection. Equations of state with a second-order phase transition like CMF might have analogous fingerprints in gravitational waves.
While gravitational wave signatures from the PCS core
are beyond the frequency range of current gravitational-wave detectors, concepts for kilohertz detectors that could probe the relevant high-frequency regime are already being explored \citep{ackley_20}.
Future simulations will need to more thoroughly investigate the potential to leverage gravitational-wave and neutrino detection to constrain the physics of the hadron-quark phase transition.

The thermodynamic conditions obtained in extreme astrophysical events like neutron stars mergers and CCSNe can be probed with present beam energies at heavy ion collision laboratories such as \texttt{GSI}~\citep{most_2022}.  Due to inverted convection, the conditions for the majority of progenitors in our CCSN simulations using the CMF EoS show high specific entropies ($s \sim 4 k_\mathrm{B}/\mathrm{baryon}$), temperatures ($T \sim 70\,\mathrm{MeV}$) and densities ($n\sim 7 \rho_\mathrm{sat}$) which exceed the conditions reached in hydrodynamic simulations of heavy-ion collisions at the \texttt{SIS18} accelerator and neutron star merger simulations that utilized the same underlying CMF model~\citep{Motornenko:2019arp,most_2022}. This indicates that heavy-ion collisions with higher beam energies than currently available at \texttt{GSI}, e.g. at the \texttt{SIS100} accelerator of the Facility for Antiproton and Ion Research, \texttt{FAIR} in Europe, which is presently under construction in Darmstadt, would be necessary in order to study conditions applicable to CCSNe.

\section*{Acknowledgements}
PJ thanks R. Mardling for fruitful discussions.
BM acknowledges support by ARC Future Fellowship FT160100035.  This work is based on simulations performed within computer time allocations from Astronomy Australia Limited's ASTAC scheme,  the National Computational Merit Allocation Scheme (NCMAS), and an Australasian Leadership Computing Grant on the NCI NF supercomputer Gadi.  This research was supported by resources provided by the Pawsey Supercomputing Centre, with funding from the Australian Government and the Government of Western Australia. 
This research was supported, in part, by the Australian Research Council (ARC) Centre of Excellence (CoE) for Gravitational Wave Discovery (OzGrav) through project number CE170100004, by the ARC CoE for All Sky Astrophysics in 3 Dimensions (ASTRO 3D) through project number CE170100013, and by the National Science Foundation under Grant No. PHY-1430152 (JINA Center for the Evolution of the Elements, JINA-CEE).
JS thanks the Samson AG for support. AM acknowledges the Stern-Gerlach Postdoctoral fellowship of the Stiftung Polytechnische Gesellschaft. HS
acknowledges the Walter Greiner Gesellschaft zur F\"orderung
der physikalischen Grundlagenforschung e.V. through the Judah M. Eisenberg Laureatus Chair at Goethe Universit\"at.

\section*{Data Availability}
The data underlying this article will be shared on reasonable request to the  authors, subject to considerations of intellectual property law.

\bibliography{bib.bib}
\label{lastpage}

\end{document}

%% file: tables/table_dd2_1.4.tex
\begin{tabular}{cccccccccccccccc}
\toprule
Metallicity &   $M$ & $M_\text{PCS}(t_\text{b,2})$ & EoS &  Explosion & $t_\mathrm{BH}$ &   $t_\text{1,b}$ &  $t_\text{MP}$ & $t_\text{2,b}$   &   $\alpha$  & $\xi_{2.5}$ &    $\mu_4$ & $M_4\mu_4$ & $C^\text{inv}_\text{L}$  \\
            & (M$_\odot$) & ($M_\odot$) & & & (s) & (s) & (s) & (s) \\
\midrule
    solar &  14 &  1.79 & DD2F\_SF &   no &            2.056 &   0.223 & 2.034  & 2.056 &    0.397 &       0.115 &   0.079 &      0.124 & yes \\
    solar &  15 &  1.87 & DD2F\_SF &   no &            1.795 &   0.240 & 1.782  & 1.795 &   0.315 &       0.165 &   0.082 &      0.141 & yes \\
    solar &  16 &  1.76 & DD2F\_SF &  yes &            -    &   0.176  & 1.971  & 2.007 &    0.488 &       0.142 &   0.073 &      0.108 & yes \\
    solar &  17 &  1.75 & DD2F\_SF &   no &            2.187 &   0.218 & 2.129 & 2.187 &   0.423 &       0.127 &   0.052 &      0.084 & yes \\
    solar &  18 &  1.95 & DD2F\_SF &   no &            1.542 &   0.252 & 1.541  & 1.542 &   0.211 &       0.248 &   0.101 &      0.184 & yes \\
    solar &  19 &  1.94 & DD2F\_SF &   no &            1.599 &   0.263 & 1.560 & 1.599 &    0.227 &       0.246 &   0.095 &      0.167 & yes \\
    solar &  20 &  1.97 & DD2F\_SF &   no &            1.516 &   0.264 & 1.515 & 1.516 &    0.258 &       0.237 &   0.103 &      0.195 & yes \\
    solar &  21 &  2.19 & DD2F\_SF &   no &            1.048 &   0.312 & 1.048 & 1.048 &    0.200 &       0.432 &   0.152 &      0.321 & yes \\
    solar &  22 &  2.15 & DD2F\_SF &   no &            1.126 &   0.307 &  1.126 & 1.126 &   0.186 &       0.402 &   0.144 &      0.295 & yes \\
    solar &  23 &  2.04 & DD2F\_SF &   no &            1.368 &   0.290 &  1.368 & 1.368 &    0.222 &       0.312 &   0.109 &      0.208 & yes \\
    solar &  24 &  1.97 & DD2F\_SF &   no &            1.560 &   0.272 & 1.368  & 1.560 &    0.174 &      0.265 &   0.092 &      0.168 & yes \\
    solar &  25 &  2.03 & DD2F\_SF &   no &            1.388 &   0.285 & 1.387  & 1.388 &   0.228 &       0.304 &   0.110 &      0.205 & yes \\
    solar &  30 &  2.21 & DD2F\_SF &   no &            1.022 &   0.322 & 1.021  & 1.022 &  0.264 &       0.474 &   0.194 &      0.405 & yes \\
    solar &  35 &  2.33 & DD2F\_SF &   no &            0.876 &   0.351 & 0.876  & 0.876 &   0.225 &       0.597 &   0.261 &      0.597 & yes \\
     \noalign{\medskip}
    primordial &  14 & 1.71 & DD2F\_SF &  yes &                 - &   0.186 &   2.220 & 2.336 &   0.513 &       0.047 &   0.044 &      0.071 & yes \\
    primordial &  15 & 1.79 &  DD2F\_SF &   failed &            - &   0.156 &  3.320 & 2.554 &   0.495 &       0.072 &   0.039 &      0.060 & yes \\
    primordial &  16 & 1.77 & DD2F\_SF &   no &             2.063 &   0.206 &  2.045 & 2.063 &   0.414 &       0.136 &   0.074 &      0.117 & yes \\
    primordial &  17 & 1.91 & DD2F\_SF &   no &             1.656 &   0.226 & 1.654  & 1.656 &    0.297 &       0.213 &   0.103 &      0.180 & yes \\
    primordial &  18 &  1.76 & DD2F\_SF &   failed &            - &   0.192 &  2.227 & 2.467 &   0.458 &       0.114 &   0.051 &      0.076 & yes \\
    primordial &  19 &  1.87 & DD2F\_SF &   failed &        1.727 &   0.214 &  1.726 & 1.727 &    0.317 &       0.197 &   0.090 &      0.147 & yes \\
    primordial &  20 &  1.84 & DD2F\_SF &   no &            1.730 &   0.192 & 1.726  & - &    0.354 &       0.170 &   0.102 &      0.150 & yes \\
    primordial &  21 &  1.86 & DD2F\_SF &   no &            1.680 &   0.198 &  1.677 & - &    0.347 &       0.202 &   0.099 &      0.153 & yes \\
    primordial &  22 &  1.83 & DD2F\_SF &   no &            1.746 &   0.182 &  1.743 & - &    0.346 &       0.206 &   0.094 &      0.142 & yes \\
    primordial &  23 &  1.86 & DD2F\_SF &   no &            1.769 &   0.224 & 1.768  & 1.769 &    0.323 &       0.190 &   0.086 &      0.140 & yes \\
    primordial &  24 &  2.16 & DD2F\_SF &   no &            1.105 &   0.308 & 1.105  & 1.105 &    0.258 &       0.413 &   0.163 &      0.325 & yes \\
    primordial &  25 &  2.21 & DD2F\_SF &   no &            0.996 &   0.304 &  0.996 & 0.996 &   0.265 &       0.446 &   0.153 &      0.332 & yes \\
   primordial&  30 &  1.95 & DD2F\_SF &   no &            1.539 &   0.244 &  1.538 & 1.539 &   0.268 &       0.237 &   0.106 &      0.185 & yes \\
   primordial&  35 &  2.25 & DD2F\_SF &   no &            0.944 &   0.320 & 0.944  & 0.944 &   0.248 &       0.487 &   0.190 &      0.407 & yes \\
   primordial&  40 &  2.37 & DD2F\_SF &   no &            0.816 &   0.333 & 0.816  & 0.816 &    0.127 &       0.645 &   0.366 &      0.764 & yes \\
   primordial&  45 &  2.36 & DD2F\_SF &   no &            0.826 &   0.333 &  0.826 & 0.826 &   0.226 &       0.635 &   0.345 &      0.740 & yes \\
   primordial&  50 &  2.28 & DD2F\_SF &   no &            0.864 &   0.332 &  0.826 & 0.864 &   0.215 &       0.593 &   0.261 &      0.584 & yes \\
   primordial&  55 &  2.34 & DD2F\_SF &   no &            0.808 &   0.305 &  0.807 & 0.808 &   0.233 &       0.635 &   0.385 &      0.679 & yes \\
   primordial&  60 &  2.30 & DD2F\_SF &   no &            0.873 &   0.299 &  0.872 & 0.873 &   0.223 &       0.585 &   0.255 &      0.489 & yes \\
   primordial&  65 &  2.32 & DD2F\_SF &   no &            0.888 &   0.316 & 0.888  & 0.888 &   0.229 &       0.587 &   0.248 &      0.490 & yes \\
   primordial&  70 &  2.40 & DD2F\_SF &   no &            0.860 &   0.350 & 0.860  & 0.860 &   0.212 &       0.639 &   0.244 &      0.533 & yes \\
   primordial&  75 &  2.44 & DD2F\_SF &   no &            0.825 &   0.352 & 0.824  & 0.825 &   0.198 &       0.678 &   0.279 &      0.601 & yes \\
   primordial&  80 &  2.48 & DD2F\_SF &   no &            0.820 &   0.378 &  0.819 & 0.820 &   0.257 &       0.711 &   0.283 &      0.640 & yes \\
   primordial&  85 &  2.64 & DD2F\_SF &   no &            0.752 &   0.412 &   0.752 & 0.752 &   0.220 &       0.872 &   0.348 &      0.844 & yes \\
   primordial&  90 &  2.67 & DD2F\_SF &   no &            0.725 &   0.416 &  0.706 & 0.725 &  0.209 &       0.922 &   0.405 &      0.981 & yes \\
   primordial&  95 &  2.63 & DD2F\_SF &   no &            0.769 &   0.433 &  0.765 & 0.769 &  0.232 &       0.875 &   0.347 &      0.854 & yes \\
\bottomrule
\end{tabular}

%% file: tables/table_stosB145.tex
\begin{tabular}{cccccccccccccccc}
\toprule
Metallicity &   $M$ & $M_\text{PCS}(t_\text{b,2})$ & EoS &  Explosion & $t_\mathrm{BH}$ &   $t_\text{1,b}$ &  $t_\text{MP}$  & $t_\text{2,b}$  &    $\alpha$  & $\xi_{2.5}$ &    $\mu_4$ & $M_4\mu_4$ & $C^\text{inv}_\text{L}$  \\
            & (M$_\odot$) & ($M_\odot$) & & & (s) & (s) & (s) &(s)\\
\midrule
    solar &  21 &  2.19 & STOS &   no &           1.037 &   0.262 &  0.356  & - &    0.176 &       0.432 &   0.152 &      0.321 & yes \\
    solar &  22 &  2.19 & STOS &   no &           1.289 &   0.258 &  0.374  & - &    0.150 &       0.402 &   0.144 &      0.295 & yes\\
    solar &  30 &  2.20 & STOS &  no &            0.961 &   0.272 &  0.390 & - &   0.173 &       0.474 &   0.194 &      0.405 & yes\\
    solar &  35 &  2.21 & STOS &  no &            0.731 &   0.296 & 0.404   & - &   0.185 &       0.597 &   0.261 &      0.597 & yes\\
    \noalign{\medskip}
    primordial &  23 &  2.26 &STOS &   no &            4.486 &   0.190 & 0.310 &    - &    0.214 &       0.190 &   0.086 &      0.140 & yes\\
    primordial &  24 &  2.19 &STOS &   no &            1.220 &   0.260 & 0.374  &  - &    0.191 &       0.413 &   0.163 &      0.325 & yes\\
    primordial &  25 &  2.20 &STOS &   no &            0.939 &   0.258 &  0.370  & - &    0.168 &       0.446 &   0.153 &      0.332  & yes\\
    primordial &  35 &  2.20 &STOS &   no &            0.842 &   0.272 &  0.372  & - &    0.153 &       0.487 &   0.190 &      0.407  & yes\\
    primordial &  40 &  2.21 &STOS &   no &            0.670 &   0.284 &  0.390  & - &     0.214 &       0.645 &   0.366 &      0.764  & yes\\
    primordial &  45 &  2.21 &STOS &   no &            0.680 &   0.284 &  0.402  & - &   0.157 &       0.635 &   0.345 &      0.740  & yes\\
    primordial &  50 &  2.21 &STOS &   no &            0.735 &   0.284 &  0.386  & - &    0.109 &       0.593 &   0.261 &      0.584  & yes\\
    primordial &  55 &  2.21 &STOS &   no &            0.694 &   0.264 & 0.356   & - &    0.187 &       0.635 &   0.385 &      0.679  & yes\\
    primordial &  60 &  2.20 &STOS &   no &            0.755 &   0.258 &  0.354  & - &    0.169 &       0.585 &   0.255 &      0.489  & yes\\
    primordial &  65 &  2.20 &STOS &   no &            0.746 &   0.268 &  0.388  & - &    0.170 &       0.587 &   0.248 &      0.490  & yes\\
    primordial &  70 &  2.25 &STOS &   failed   &  0.658 &   0.304 &       0.402   &   0.658 &     0.177 &       0.639 &   0.244 &      0.533  & yes\\
    primordial &  75 &  2.28 &STOS &   failed &  0.650 &   0.306 & 0.406 &  0.650 &    0.133 &       0.678 &   0.279 &      0.601  & yes\\
    primordial &  80 &  2.33 &STOS &   no &            0.642 &    0.330 & 0.422 & - &    0.198 &       0.711 &   0.283 &      0.640 & yes\\
    primordial &  85 &  2.48 &STOS &   no &            0.612   &0.362 & 0.452 &- & 0.181 &       0.872 &   0.348 &      0.844  & yes\\
    primordial &  90 &  2.52 &STOS &   no &            0.607  &0.370 &   0.470 & - &    0.178 &       0.922 &   0.405 &      0.981  & yes\\
    primordial &  95 &  2.51 &STOS &   no &            0.620 &  0.378 & 0.480 & -  & 0.193 &       0.875 &   0.347 &      0.854  & yes\\
    primordial &  100 &  2.21 &STOS &  no &            0.731 &   0.296 & 0.418 &-  & 0.172 &       0.597 &   0.261 &      0.597  & yes\\
\bottomrule
\end{tabular}

%% file: tables/table_cmf.tex
\begin{tabular}{ccccccccccccccc}
\toprule
Metallicity &   $M$ & $M_\text{PCS}(t_\text{b,2})$ & EoS &  Explosion & $t_\mathrm{BH}$ &   $t_\text{1,b}$ &  $t_\text{2,b}$    &   $\alpha$  & $\xi_{2.5}$ &    $\mu_4$ & $M_4\mu_4$ & $C^\text{inv}_\text{L}$  \\
            & (M$_\odot$) & ($M_\odot$) & & & (s) & (s) & (s)\\
\midrule
    solar &  16 &  2.22 & CMF &   no &            - &   0.196 &   - &    0.529 &       0.142 &   0.073 &      0.108 & yes\\
    solar &  18 & 2.49 & CMF &   no &            4.641 &   0.272 &   - &    0.229 &       0.248 &   0.101 &      0.184 & yes\\
    solar &  19 &  2.48 & CMF &   no &            4.683 &   0.282 &   - &    0.230 &       0.246 &   0.095 &      0.167 & yes\\
    solar &  20 &  2.49 & CMF &   no &            5.345 &   0.282 &   - &    0.185 &       0.237 &   0.103 &      0.195 & yes\\
    solar &  21 &  2.41 & CMF &   no &            1.885 &   0.326 &   - &    0.177 &       0.432 &   0.152 &      0.321 & yes \\
    solar &  22 &  2.41 & CMF &   no &            2.213 &   0.322 &   - &    0.221 &       0.402 &   0.144 &      0.295 & yes \\
    solar &  23 &  2.45 & CMF &   no &            3.472 &   0.312 &   - &  0.203 &       0.312 &   0.109 &      0.208 & yes \\ 
    solar &  24 &  2.48 & CMF &   no &            4.450 &   0.294 &   - &   0.188 &       0.265 &   0.092 &      0.168 & yes\\
    solar &  25 &  2.45 & CMF &   no &            3.481 &   0.306 &   - &    0.185 &       0.304 &   0.110 &      0.205 & yes \\
    solar &  30 &  2.40 & CMF &   no &            1.868 &   0.336 &   - &    0.201 &       0.474 &   0.194 &      0.405 & yes \\
    \noalign{\medskip}
    primordial &  14 &  1.90 & CMF &   no &            - &   0.208 &   - &   0.602 &       0.047 &   0.044 &      0.071 & yes \\
    primordial &  15 &  1.83 & CMF &   no &            - &   0.176 &   - &    0.616 &       0.072 &   0.039 &      0.060 & yes \\
    primordial &  16 &  2.10 & CMF &   no &            - &   0.228 &   - &    0.559 &       0.136 &   0.074 &      0.117 & yes \\
    primordial &  17 &  2.40 & CMF &   no &            - &   0.250 &   - &    0.474 &       0.213 &   0.103 &      0.180 & yes \\
    primordial &  18 &  1.99 & CMF &   no &            - &   0.212 &   - &    0.582 &       0.114 &   0.051 &      0.076 & yes \\
    primordial &  19 &  2.45 & CMF &   no &            - &   0.238 &   - &    0.447 &       0.197 &   0.090 &      0.147 & yes \\
    primordial &  20 &  2.36 & CMF &   no &            - &   0.214 &   - &    0.487 &       0.170 &   0.102 &      0.150 & yes \\
    primordial &  21 &  2.51 & CMF &   no &            5.861 &   0.220 &   - &    0.217 &       0.202 &   0.099 &      0.153 & yes \\
    primordial &  22 &  2.51 & CMF &   no &            5.978 &   0.202 &   - &   0.200 &       0.206 &   0.094 &      0.142 & yes \\
    primordial &  23 &  2.39 & CMF &   no &            5.899 &   0.246 &   - &    0.477 &       0.190 &   0.086 &      0.140 & yes \\
    primordial &  24 &  2.41 & CMF &   no &            2.256 &   0.326 &   - &    0.209 &       0.413 &   0.163 &      0.325 & yes \\
    primordial &  25 &  2.41 & CMF &   no &            1.722 &   0.322 &   - &    0.183 &       0.446 &   0.153 &      0.332 & yes\\
    primordial &  30 &  2.49 & CMF &   no &            5.034 &   0.264 &   - &    0.230 &       0.237 &   0.106 &      0.185 & yes\\
    primordial &  35 &  2.42 & CMF &   no &            1.545 &   0.338 &   - &    0.188 &       0.487 &   0.190 &      0.407 & yes \\
    primordial &  40 &  2.49 & CMF &   no &            0.960 &   0.348 &   - &    0.125 &       0.645 &   0.366 &      0.764 & yes \\
    primordial &  45 &  2.48 & CMF &   no &            0.986 &   0.348 &   - &    0.077 &       0.635 &   0.345 &      0.740 & yes\\
    primordial &  50 &  2.47 & CMF &   no &            1.082 &   0.348 &   - &    0.181 &       0.593 &   0.261 &      0.584 & yes \\
    primordial &  55 &  2.49 & CMF &   no &            0.959 &   0.328 &   - &    0.148 &       0.635 &   0.385 &      0.679 & yes\\
    primordial &  60 &  2.46 & CMF &   no &            1.073 &   0.322 &   - &    0.131 &       0.585 &   0.255 &      0.489 & yes \\
    primordial &  65 &  2.46 & CMF &   no &            1.091 &   0.334 &   - &    0.016 &       0.587 &   0.248 &      0.490 & yes\\
    primordial &  70 &  2.49 & CMF &   no &            1.009 &   0.366 &   - &    0.192 &       0.639 &   0.244 &      0.533 & yes\\
    primordial &  75 &  2.51 & CMF &   no &            0.933 &   0.370 &   - &    0.190 &       0.678 &   0.279 &      0.601 & yes \\
    primordial &  80 &  2.54 & CMF &   no &            0.900 &   0.392 &   - &   0.175 &       0.711 &   0.283 &      0.640 & yes \\
    primordial &  85 &  2.59 & CMF &   no &            0.744 &   0.424 &   - &    0.180 &       0.872 &   0.348 &      0.844 & yes \\
    primordial &  95 &  2.60 & CMF &   no &            0.747 &   0.440 &   - &    0.210 &       0.875 &   0.347 &      0.854 & yes\\
    primordial &  100 &  2.41 & CMF &   no &            1.668 &   0.308 &   - &   0.216 &       0.404 &   0.176 &      0.349 & yes\\
\bottomrule
\end{tabular}

%% file: tables/lepp.tex
\begin{tabular}{cll|clll}
\toprule
LEPP                 & \texttt{z14}       & \texttt{s16}        &  Iron           & \texttt{z14}        & \texttt{s16}        \\
                 &     &       &  group           &         &        \\
\midrule
${}^{78}\mathrm{Kr}$ & 2.145e-07 & 2.003e-08 & ${}^{55}\mathrm{Mn}$ & 1.649e-05 & 2.780e-04 \\
${}^{80}\mathrm{Kr}$ & 5.820e-07 & 1.109e-07 & ${}^{54}\mathrm{Fe}$ & 6.961e-05 & 2.313e-03 \\
${}^{82}\mathrm{Kr}$ & 1.883e-07 & 5.005e-07 & ${}^{56}\mathrm{Fe}$ & 3.756e-03 & 2.727e-02 \\
${}^{83}\mathrm{Kr}$ & 1.942e-06 & 1.794e-06 & ${}^{57}\mathrm{Fe}$ & 2.090e-04 & 1.351e-03 \\
${}^{84}\mathrm{Kr}$ & 1.261e-05 & 1.299e-05 & ${}^{58}\mathrm{Fe}$ & 1.926e-06 & 5.601e-04 \\
${}^{86}\mathrm{Kr}$ & 2.035e-05 & 6.922e-06 & ${}^{59}\mathrm{Co}$ & 1.916e-05 & 3.110e-04 \\
${}^{85}\mathrm{Rb}$ & 2.079e-06 & 2.133e-06 & ${}^{58}\mathrm{Ni}$ & 2.424e-04 & 1.793e-03 \\
${}^{87}\mathrm{Rb}$ & 3.526e-06 & 8.081e-07 & ${}^{60}\mathrm{Ni}$ & 2.472e-04 & 1.174e-03 \\
${}^{84}\mathrm{Sr}$ & 5.079e-08 & 2.479e-08 & ${}^{61}\mathrm{Ni}$ & 2.532e-05 & 1.489e-04 \\
${}^{86}\mathrm{Sr}$ & 1.194e-07 & 1.325e-07 & ${}^{62}\mathrm{Ni}$ & 1.434e-04 & 4.197e-04 \\
${}^{87}\mathrm{Sr}$ & 3.426e-08 & 8.903e-08 & ${}^{64}\mathrm{Ni}$ & 4.831e-05 & 1.068e-04 \\
${}^{88}\mathrm{Sr}$ & 5.524e-05 & 8.335e-06 & ${}^{63}\mathrm{Cu}$ & 1.366e-05 & 1.309e-04 \\
${}^{89}\mathrm{Y}$  & 5.422e-06 & 1.109e-06 & ${}^{65}\mathrm{Cu}$ & 9.219e-06 & 2.692e-05 \\
${}^{90}\mathrm{Zr}$ & 3.086e-05 & 3.000e-06 & ${}^{64}\mathrm{Zn}$ & 8.314e-05 & 1.071e-04 \\
${}^{91}\mathrm{Zr}$ & 1.118e-06 & 7.637e-07 & ${}^{66}\mathrm{Zn}$ & 5.407e-05 & 3.851e-05 \\
${}^{92}\mathrm{Zr}$ & 3.449e-07 & 1.775e-06 & ${}^{67}\mathrm{Zn}$ & 4.827e-06 & 7.280e-06 \\
${}^{94}\mathrm{Zr}$ & 2.487e-07 & 2.349e-06 & ${}^{68}\mathrm{Zn}$ & 1.996e-05 & 3.832e-05 \\
${}^{96}\mathrm{Zr}$ & 1.693e-08 & 2.415e-07 & ${}^{70}\mathrm{Zn}$ & 6.634e-07 & 8.277e-07 \\
${}^{93}\mathrm{Nb}$ & 8.254e-08 & 4.490e-07 &                      &           &           \\
                     &           &           &                      &           &    \\      
\bottomrule
\end{tabular}